\documentclass[twocolumn,twocolappendix]{aastex63}
\hypersetup{linkcolor=blue,citecolor=blue,filecolor=blue,urlcolor=blue}
\usepackage{amsmath}

\newcommand{\tess}[1]{\emph{TESS}#1}

\newcommand{\teff}[1]{$T_{\text{eff}}$#1}
\newcommand{\prot}[1]{$P_{\text{rot}}$#1}
\newcommand{\logrhk}[1]{$\log{R'_{\text{HK}}}$#1}
\newcommand{\logg}[1]{$\log{g}$#1}
\newcommand{\vsini}[1]{$v\sin{i}$#1}

\newcommand{\mps}[1]{m s$^{-1}$#1}

\newcommand{\Rearth}[1]{$R_{\oplus}$#1}
\newcommand{\Rsun}[1]{$R_{\odot}$#1}
\newcommand{\Mearth}[1]{$M_{\oplus}$#1}
\newcommand{\Msun}[1]{$M_{\odot}$#1}
\newcommand{\name}[1]{TOI-1634#1}
\newcommand{\mplanet}[1]{$4.91^{+0.68}_{-0.70}$#1}
\newcommand{\rplanet}[1]{$1.790^{+0.080}_{-0.081}$#1}

\shortauthors{Cloutier et al.}
\shorttitle{}

\begin{document}
\title{\name{} b: an Ultra-Short Period Keystone Planet Sitting Inside the M Dwarf Radius Valley}

\suppressAffiliations

\author[0000-0001-5383-9393]{Ryan Cloutier}
\altaffiliation{Banting Fellow}
\affiliation{Center for Astrophysics $\vert$ Harvard \& Smithsonian, 60 Garden
  Street, Cambridge, MA, 02138, USA}

\author[0000-0002-9003-484X]{David Charbonneau}  
\affiliation{Center for Astrophysics $\vert$ Harvard \& Smithsonian, 60 Garden
  Street, Cambridge, MA, 02138, USA}


\author[0000-0002-3481-9052]{Keivan G. Stassun}  
\affiliation{Department of Physics \& Astronomy, Vanderbilt University,
6301 Stevenson Center Lane, Nashville, TN, 37235, USA}

\author{Felipe Murgas}  
\affiliation{Instituto de Astrofísica de Canarias, C/ Vía Láctea s/n, E-38205 La Laguna, Spain}

\author[0000-0001-7254-4363]{Annelies Mortier}  
\affiliation{Astrophysics Group, Cavendish Laboratory, University of Cambridge,
  J.J. Thomson Avenue, Cambridge CB3 0HE, UK}

\author[0000-0001-8879-7138]{Robert Massey}  
\affiliation{American Association of Variable Star Observers (AAVSO), 49 Bay
State Rd, Cambridge, MA, 02138, USA}

\author[0000-0001-6513-1659]{Jack J. Lissauer}  
\affiliation{NASA Ames Research Center, Moffett Field, CA, 94035, USA}

\author{David W. Latham}  
\affiliation{Center for Astrophysics $\vert$ Harvard \& Smithsonian, 60 Garden
  Street, Cambridge, MA, 02138, USA}

\author{Jonathan Irwin}  
\affiliation{Center for Astrophysics $\vert$ Harvard \& Smithsonian, 60 Garden
  Street, Cambridge, MA, 02138, USA}

\author{Rapha\"{e}lle D. Haywood}  
\affiliation{Astrophysics Group, University of Exeter, Exeter EX4 2QL, UK}

\author[0000-0002-4308-2339]{Pere Guerra}  
\affiliation{Observatori Astron\`omic Albany\`a, Cam\'i de Bassegoda S/N, 
  Albany\`a 17733, Girona, Spain}

\author[0000-0002-5443-3640]{Eric Girardin}  
\affiliation{Grand Pra Observatory, 1984 Les Hauderes, Switzerland}

\author[0000-0002-8965-3969]{Steven A. Giacalone}  
\affiliation{Department of Astronomy, University of California, Berkeley, 
  Berkeley, CA 94720, USA}

\author{Pau Bosch-Cabot}  
\affiliation{Observatori Astron\`omic Albany\`a, Cam\'i de Bassegoda S/N,
  Albany\`a 17733, Girona, Spain}

\author[0000-0001-6637-5401]{Allyson Bieryla}  
\affiliation{Center for Astrophysics $\vert$ Harvard \& Smithsonian, 60 Garden
  Street, Cambridge, MA, 02138, USA}


\author[0000-0002-4265-047X]{Joshua Winn} 
\affiliation{Department of Astrophysical Sciences, Princeton University,
Princeton, NJ 08544, USA}

\author[0000-0002-9718-3266]{Christopher A. Watson}  
\affiliation{Astrophysics Research Centre, School of Mathematics and Physics,
Queen's University Belfast, Belfast, BT7 1NN, UK}

\author[0000-0001-6763-6562]{Roland Vanderspek}  
\affiliation{Department of Earth, Atmospheric and Planetary Sciences, and 
Kavli Institute for Astrophysics and Space Research, Massachusetts Institute 
of Technology, Cambridge, MA 02139, USA}

\author{St\'ephane Udry}  
\affiliation{Observatoire Astronomique de l'Universit\'e de Gen\`eve, 51 chemin
  des Maillettes, 1290 Versoix, Switzerland}

\author[0000-0002-6510-0681]{Motohide Tamura}  
\affiliation{Dept. of Astronomy, Graduate School of Science, The University of Tokyo, 7-3-1 Hongo, Bunkyo-ku, Tokyo 113-0033, Japan}
\affiliation{Astrobiology Center, National Institutes of Natural Sciences, 2-21-1 Osawa, Mitaka, Tokyo 181-8588, Japan}
\affiliation{National Astronomical Observatory of Japan, 2-21-1 Osawa, Mitaka, Tokyo 181-8588, Japan}

\author[0000-0002-7504-365X]{Alessandro Sozzetti}  
\affiliation{INAF - Osservatorio Astrofisico di Torino, Strada Osservatorio 20,
  Pino Torinese (To) 10025, Italy}

\author{Avi Shporer}  
\affiliation{Department of Physics and Kavli Institute for Astrophysics and
Space Research, Massachusetts Institute of Technology, Cambridge, MA 02139, USA}

\author{Damien S\'egransan}  
\affiliation{Observatoire Astronomique de l'Universit\'e de Gen\`eve, 51 chemin
  des Maillettes, 1290 Versoix, Switzerland}

\author[0000-0002-6892-6948]{Sara Seager}  
\affiliation{Department of Physics and Kavli Institute for Astrophysics and
Space Research, Massachusetts Institute of Technology, Cambridge, MA 02139, USA}
\affiliation{Department of Earth, Atmospheric and Planetary Sciences,
Massachusetts Institute of Technology, Cambridge, MA 02139, USA}
\affiliation{Department of Aeronautics and Astronautics, MIT, 77 Massachusetts
Avenue, Cambridge, MA 02139, USA}

\author[0000-0002-2454-768X]{Arjun B. Savel}  
\affiliation{Department of Astronomy, University of Maryland, College Park, MD}
\affiliation{Department of Astronomy, University of California, Berkeley,
  Berkeley, CA 94720, USA}

\author[0000-0001-7014-1771]{Dimitar Sasselov}  
\affiliation{Center for Astrophysics $\vert$ Harvard \& Smithsonian, 60 Garden
  Street, Cambridge, MA, 02138, USA}

\author[0000-0003-4724-745X]{Mark Rose}  
\affiliation{NASA Ames Research Center, Moffett Field, CA, 94035, USA}

\author[0000-0003-2058-6662]{George Ricker}  
\affiliation{Department of Earth, Atmospheric and Planetary Sciences, and 
Kavli Institute for Astrophysics and Space Research, Massachusetts Institute 
of Technology, Cambridge, MA 02139, USA}

\author[0000-0002-6379-9185]{Ken Rice}  
\affiliation{SUPA, Institute for Astronomy, University of Edinburgh, Blackford
  Hill, Edinburgh, EH9 3HJ, Scotland, UK}
\affiliation{Centre for Exoplanet Science, University of Edinburgh, Edinburgh, UK}

\author[0000-0003-1309-2904]{Elisa V. Quintana} 
\affiliation{NASA Goddard Space Flight Center, 8800 Greenbelt Road, Greenbelt, MD 20771, USA}

\author[0000-0002-8964-8377]{Samuel N. Quinn}  
\affiliation{Center for Astrophysics $\vert$ Harvard \& Smithsonian, 60 Garden
  Street, Cambridge, MA, 02138, USA}

\author{Giampaolo Piotto}  
\affiliation{Dip. di Fisica e Astronomia Galileo Galilei - Universit\`a di
Padova, Vicolo dell'Osservatorio 2, 35122, Padova, Italy}

\author{David Phillips}  
\affiliation{Center for Astrophysics $\vert$ Harvard \& Smithsonian, 60 Garden
  Street, Cambridge, MA, 02138, USA}

\author{Francesco Pepe}  
\affiliation{Observatoire Astronomique de l'Universit\'e de Gen\`eve, 51 chemin
  des Maillettes, 1290 Versoix, Switzerland}

\author[0000-0002-5752-6260]{Marco Pedani} 
\affiliation{Fundaci\'on Galileo Galilei-INAF, Rambla Jos\'e Ana Fernandez
P\'erez 7, 38712 Bre\~{n}a Baja, TF, Spain}

\author[0000-0001-5519-1391]{Hannu Parviainen} 
\affiliation{Instituto de Astrofísica de Canarias, C/ Vía Láctea s/n, E-38205 La Laguna, Spain}
\affiliation{Departamento de Astrofísica, Universidad de La Laguna, E-38206 La Laguna, Spain}

\author{Enric Palle}  
\affiliation{Instituto de Astrofísica de Canarias, C/ Vía Láctea s/n, E-38205 La Laguna, Spain}
\affiliation{Departamento de Astrofísica, Universidad de La Laguna, E-38206 La Laguna, Spain}

\author[0000-0001-8511-2981]{Norio Narita}  
\affiliation{Komaba Institute for Science, The University of Tokyo, 3-8-1 Komaba, Meguro, Tokyo 153-8902, Japan}
\affiliation{JST, PRESTO, 3-8-1 Komaba, Meguro, Tokyo 153-8902, Japan}
\affiliation{Astrobiology Center, National Institutes of Natural Sciences, 2-21-1 Osawa, Mitaka, Tokyo 181-8588, Japan}
\affiliation{Instituto de Astrof\'{i}sica de Canarias (IAC), 38205 La Laguna, Tenerife, Spain}

\author[0000-0002-1742-7735]{Emilio Molinari}  
\affiliation{INAF - Osservatorio Astronomico di Cagliari, via della Scienza 5,
09047, Selargius, Italy}

\author[0000-0002-9900-4751]{Giuseppina Micela}  
\affiliation{INAF - Osservatorio Astronomico di Palermo, Piazza del Parlamento
1, I-90134 Palermo, Italy}

\author{Scott McDermott}  
\affiliation{Proto-Logic LLC, 1718 Euclid Street NW, Washington, DC 20009, USA}

\author{Michel Mayor}  
\affiliation{Observatoire Astronomique de l'Universit\'e de Gen\`eve, 51 chemin
  des Maillettes, 1290 Versoix, Switzerland}

\author[0000-0001-7233-7508]{Rachel A. Matson}   
\affiliation{U.S. Naval Observatory, Washington, DC 20392, USA}

\author{Aldo F. Martinez Fiorenzano}  
\affiliation{Fundaci\'on Galileo Galilei-INAF, Rambla Jos\'e Ana Fernandez
P\'erez 7, 38712 Bre\~{n}a Baja, TF, Spain}

\author{Christophe Lovis}  
\affiliation{Observatoire Astronomique de l'Universit\'e de Gen\`eve, 51 chemin
  des Maillettes, 1290 Versoix, Switzerland}

\author[0000-0003-3204-8183]{Mercedes L\'opez-Morales}  
\affiliation{Center for Astrophysics $\vert$ Harvard \& Smithsonian, 60 Garden
  Street, Cambridge, MA, 02138, USA}

\author[0000-0001-9194-1268]{Nobuhiko Kusakabe}  
\affiliation{Astrobiology Center, National Institutes of Natural Sciences, 2-21-1 Osawa, Mitaka, Tokyo 181-8588, Japan}
\affiliation{National Astronomical Observatory of Japan, 2-21-1 Osawa, Mitaka, Tokyo 181-8588, Japan}

\author[0000-0002-4625-7333]{Eric L. N. Jensen}  
\affiliation{Dept. of Physics \& Astronomy, Swarthmore College, Swarthmore PA
  19081, USA}

\author[0000-0002-4715-9460]{Jon M. Jenkins}  
\affiliation{NASA Ames Research Center, Moffett Field, CA, 94035, USA}

\author{Chelsea X. Huang} 
\affiliation{Department of Physics and Kavli Institute for Astrophysics and
Space Research, Massachusetts Institute of Technology, Cambridge, MA 02139, USA}

\author[0000-0002-2532-2853]{Steve B. Howell}  
\affiliation{NASA Ames Research Center, Moffett Field, CA, 94035, USA}

\author{Avet Harutyunyan} 
\affiliation{Fundaci\'on Galileo Galilei-INAF, Rambla Jos\'e Ana Fernandez
P\'erez 7, 38712 Bre\~{n}a Baja, TF, Spain}

\author{G{\'a}bor F{\H u}r{\' e}sz} 
\affiliation{Department of Physics and Kavli Institute for Astrophysics and
Space Research, Massachusetts Institute of Technology, Cambridge, MA 02139, USA}

\author[0000-0002-4909-5763]{Akihiko Fukui}  
\affiliation{Dept. of Earth and Planetary Science, Graduate School of Science, The University of Tokyo, 7-3-1 Hongo, Bunkyo-ku, Tokyo 113-0033, Japan}
\affiliation{Instituto de Astrof\'isica de Canarias, V\'ia L\'actea s/n, E-38205 La Laguna, Tenerife, Spain}

\author[0000-0002-9789-5474]{Gilbert A. Esquerdo}  
\affiliation{Center for Astrophysics $\vert$ Harvard \& Smithsonian, 60 Garden
  Street, Cambridge, MA, 02138, USA}

\author[0000-0002-2341-3233]{Emma Esparza-Borges}  
\affiliation{Departamento de Astrofísica, Universidad de La Laguna, E-38206 La Laguna, Spain}

\author{Xavier Dumusque}  
\affiliation{Observatoire Astronomique de l'Universit\'e de Gen\`eve, 51 chemin
  des Maillettes, 1290 Versoix, Switzerland}

\author[0000-0001-8189-0233]{Courtney D. Dressing}  
\affiliation{Department of Astronomy, University of California, Berkeley, 
  Berkeley, CA 94720, USA}

\author{Luca Di Fabrizio} 
\affiliation{Fundaci\'on Galileo Galilei-INAF, Rambla Jos\'e Ana Fernandez
P\'erez 7, 38712 Bre\~{n}a Baja, TF, Spain}

\author[0000-0001-6588-9574]{Karen A. Collins}  
\affiliation{Center for Astrophysics $\vert$ Harvard \& Smithsonian, 60 Garden
  Street, Cambridge, MA, 02138, USA}

\author{Andrew Collier Cameron}  
\affiliation{School of Physics and Astronomy, University of St Andrews, North
Haugh, St Andrews, Fife, KY16 9SS, UK}

\author[0000-0002-8035-4778]{Jessie L. Christiansen}  
\affiliation{Caltech/IPAC, 1200 E. California Blvd. Pasadena, CA 91125, USA}

\author{Massimo Cecconi}  
\affiliation{Fundaci\'on Galileo Galilei-INAF, Rambla Jos\'e Ana Fernandez
P\'erez 7, 38712 Bre\~{n}a Baja, TF, Spain}

\author{Lars A. Buchhave}  
\affiliation{DTU Space, National Space Institute, Technical University of 
Denmark, Elektrovej 328, DK-2800 Kgs. Lyngby, Denmark}

\author[0000-0001-9978-9109]{Walter Boschin} 
\affiliation{Fundaci\'on Galileo Galilei-INAF, Rambla Jos\'e Ana Fernandez
P\'erez 7, 38712 Bre\~{n}a Baja, TF, Spain}
\affiliation{Instituto de Astrofísica de Canarias, C/ Vía Láctea s/n, E-38205 La Laguna, Spain}
\affiliation{Departamento de Astrofísica, Universidad de La Laguna, E-38206 La Laguna, Spain}

\author[0000-0001-5125-6397]{Gloria Andreuzzi} 
\affiliation{INAF - Osservatorio Astronomico di Roma Via Frascati 33, 00078 Monte Porzio Catone (Roma) Italy}
\affiliation{Fundaci\'on Galileo Galilei-INAF, Rambla Jos\'e Ana Fernandez
P\'erez 7, 38712 Bre\~{n}a Baja, TF, Spain}

\correspondingauthor{Ryan Cloutier}
\email{ryan.cloutier@cfa.harvard.edu}

\begin{abstract}
  Studies of close-in planets orbiting M dwarfs have suggested that the
  M dwarf radius valley may
  be well-explained by distinct formation timescales between enveloped
  terrestrials, and rocky planets that form at late times in a gas-depleted
  environment. This scenario is at odds with the picture that close-in rocky
  planets form with a primordial gaseous envelope that is subsequently stripped
  away by some thermally-driven mass loss process. 
  These two physical scenarios make unique predictions of the rocky/enveloped
  transition's dependence on orbital separation such that
  studying the compositions of planets within the M dwarf radius valley
  may be able to establish the dominant physics. Here, we present the
  discovery of one such keystone planet: the ultra-short period planet
  \name{} b ($P=0.989$ days, $F=121 F_{\oplus}$, $r_p=$ \rplanet{} \Rearth{)} 
  orbiting a nearby M2 dwarf ($K_s=8.7$, $R_s=0.450$ \Rsun{,} $M_s=0.502$
  \Msun{)} and whose size and orbital period sit within the M dwarf radius
  valley. We confirm the TESS-discovered planet candidate using extensive
  ground-based follow-up campaigns, including a set of 32 precise radial
  velocity measurements from HARPS-N. We measure a planetary mass of
  \mplanet{} \Mearth{,} which makes \name{} b 
  inconsistent with an Earth-like composition at $5.9\sigma$ and thus 
  requires either an extended gaseous envelope, a large volatile-rich layer, 
  or a rocky composition that is not dominated by iron and silicates 
  to explain its mass and radius. The discovery that the bulk
  composition of \name{} b is inconsistent with that of the Earth
  supports the gas-depleted formation mechanism to explain the
  emergence of the radius valley around M dwarfs with $M_s\lesssim 0.5$ \Msun{.}
\end{abstract}


\section{Introduction}
Early-to-mid M dwarfs experience extended pre-main sequence lifetimes in
which they remain XUV active for hundreds of Myr up to about a Gyr
\citep{shkolnik14,france16}.
This does not bode well for the survival of primordial H/He envelopes around
close-in planets due to the envelope's susceptibility to hydrodynamic escape
driven by photoevaporation
\citep[e.g.][]{owen13,jin14,lopez14,chen16,jin18} or by internal heating
\citep[i.e. core-powered mass loss][]{ginzburg18,gupta19}. In such
scenarios, the largest rocky planets without envelopes increases toward
greater insolation since planets need to be more massive to retain their
envelopes. However, occurrence rate studies of close-in M dwarf
planets have revealed evidence that thermally-driven mass loss does not sculpt
the close-in M dwarf planet population \citep{cloutier20} and instead, close-in
gas-enveloped terrestrials and rocky planets formed on distinct timescales
with the latter forming at late times in a nearly gas-depleted environment
\citep{lopez18}. In this scenario, a natural outcome of terrestrial planet
formation posits that the maximum radius of rocky planets increases toward
lower insolation, in opposition to predictions from thermally-driven mass loss.
Because the thermally-driven mass loss and gas-depleted formation models make
unique predictions regarding the location of the M dwarf radius valley as a
function of insolation or period, studying the bulk compositions of planets
within the radius valley may be able to establish the dominant physics that
sculpts the close-in planet population around M dwarfs.

Since its science operations began in July 2018, NASA's Transiting
Exoplanet Survey Satellite \citep[TESS;][]{ricker15} has uncovered a wealth
of transiting planet candidates whose orbital periods and radii lie within the
radius valley, including three planets transiting early M dwarfs
(TOI-1235 b; \citealt{cloutier20c,bluhm20}, TOI-776 b; \citealt{luque21},
TOI-1685 b; \citealt{bluhm21}).
Radius valley planets whose periods $P$ and radii $r_p$ satisfy

\begin{align}
  0.11\: \log_{10}{\left( \frac{P}{\text{days}} \right)} + 1.52 \leq \frac{r_p}{R_{\oplus}}& \\ \leq -0.48\: \log_{10}{\left( \frac{P}{\text{days}} \right)} + 2.32 &,
\end{align}

\noindent \citep{cloutier20}, we refer to as keystone planets and are
valuable targets to conduct tests of the competing radius valley emergence
models across a range of stellar masses. Doing so requires that we characterize
the bulk compositions of a sample of keystone planets using precise
radial velocity measurements. Here we present the confirmation and
characterization of one such
keystone planet from TESS: \name{} b. Our study focuses on
planet validation, including the recovery of its mass, and
the implications that our results have on the emergence of the radius valley
around early M dwarfs.

In Section~\ref{sect:star} we present the properties of the host star \name{.}
In Section~\ref{sect:observations} we present the \tess{} light curve and our
suite of follow-up observations, which we use to validate the planetary nature
of the planet candidate. In Section~\ref{sect:analysis} we present our global data
analysis and its results. We conclude with a discussion and a summary of our
findings in Sects.~\ref{sect:discussion} and~\ref{sect:summary}.

\section{Stellar Characterization} \label{sect:star}
Table~\ref{tab:star} reports our adopted stellar parameters.

\begin{deluxetable}{lcc}
\tabletypesize{\small}
\tablecaption{\name{} stellar parameters.\label{tab:star}}
\tablewidth{0pt}
\tablehead{\colhead{Parameter} & \colhead{Value} & \colhead{Refs}}
\startdata 
\multicolumn{3}{c}{\emph{TOI-1634, TIC 201186294, 2MASS J03453363+3706438,}} \\
\multicolumn{3}{c}{\emph{Gaia DR3 223158499179138432}} \\
\multicolumn{3}{c}{\emph{Astrometry}} \\
Right ascension (J2015.5), $\alpha$ & 03:45:33.75 & 1,2 \\
Declination (J2015.5), $\delta$ & +37:06:44.21 & 1,2 \\
RA proper motion, $\mu_{\alpha}$ [mas yr$^{-1}$] & $81.35\pm 0.02$ & 1,2 \\
Dec proper motion, $\mu_{\delta}$ [mas yr$^{-1}$] & $13.55\pm 0.02$ & 1,2 \\
Parallax, $\pi$ [mas] & $28.512\pm 0.018$ & 1,2 \\
Distance, $d$ [pc] & $35.274\pm 0.053$ & 3 \\
\multicolumn{3}{c}{\emph{(Uncontaminated) Photometry}} \\
$V$ & $13.24\pm 0.04$ & 4 \\
$G_{\text{BP}}$ & $13.5039\pm 0.0011$ & 1,6 \\
$G$ & $12.1863\pm 0.0003$ & 1,6 \\
$G_{\text{RP}}$ & $11.0447\pm 0.0005$ & 1,6 \\
$T$ & $11.0136\pm 0.0073$ & 7 \\
$J$ & $9.564\pm 0.021$ & 4 \\
$H$ & $8.940\pm 0.021$ & 4 \\
$K_s$ & $8.699\pm 0.014$ & 4 \\
$W1$ & $8.429\pm 0.022$ & 5 \\
$W2$ & $8.325\pm 0.020$ & 5 \\
$W3$ & $8.250\pm 0.023$ & 5 \\
$W4$ & $8.266\pm 0.300$ & 5 \\
\multicolumn{3}{c}{\emph{Stellar parameters}} \\
Spectral type & M2 & 4 \\
$M_{K_s}$ & $5.88\pm 0.01$ & 4 \\
Effective temperature, \teff{} [K] & $3550\pm 69$ & 4 \\
Surface gravity, \logg{} [dex] & $4.833\pm 0.028$ & 4 \\
Metallicity, [Fe/H] [dex] & $0.23^{+0.07}_{-0.08}$ & 4 \\  
Stellar radius, $R_s$ [$R_{\odot}$] & $0.450\pm 0.013$ & 4 \\ 
Stellar mass, $M_s$ [$M_{\odot}$] & $0.502\pm 0.014$ & 4 \\
Stellar density, $\rho_s$ [g cm$^{-3}$] & $7.77^{+0.72}_{-0.62}$ & 4 \\
Stellar luminosity, $L_s$ [L$_{\odot}$] & $0.0289^{+0.0028}_{-0.0026}$ & 4 \\
\vspace{-0.15cm} Projected rotation velocity, && \\ \vspace{-0.25cm}
& $<1.3$\tablenotemark{a} & 4 \\
\vsini{} [km s$^{-1}$] && \\
\logrhk{} & $-5.39\pm 0.19$ & 4 \\
Rotation period, \prot{} [days]\tablenotemark{b} & $77^{+26}_{-20}$ & 4 \\
\enddata
\tablecomments{\textbf{References:}
  1) \citealt{gaia20}
  2) \citealt{lindegren20}
  3) \citealt{bailerjones18}
  4) this work
  5) \citealt{cutri13}
  6) \citealt{riello20}
  7) \citealt{stassun19}.}
\tablenotetext{a}{Based on the upper limits on rotational broadening from the cross-correlation function of our HARPS-N spectra.}
\tablenotetext{b}{We do not measure the stellar rotation period. Rather, \prot{} is estimated from the rotation-activity relation of \cite{astudillodefru17b}.}
\end{deluxetable}

\name{} (TIC 201186294, 2MASS J03453363+3706438, Gaia DR3 223158499179138432)
is an M2 dwarf \citep{pecaut13} at a distance of
$35.274\pm 0.053$ pc \citep{bailerjones18,gaia20,lindegren20}.
The value of the Gaia EDR3 RUWE (re-normalized unit weight error)
astrometric quality indicator reveals that \name{'s} astrometric solution
shows a large excess of 0.121 mas.\footnote{RUWE=1.23 where RUWE=1 is
  assigned to well-behaved single star solutions and RUWE$>1.4$ likely
  indicates a non-single star.} This may be indicative of a long-period
companion to \name{,} which we will revisit with our follow-up observations
in Sections~\ref{sect:sg3} and~\ref{sect:harpsn}. 
Gaia EDR3 also revealed a faint ($\Delta G=3.40$ mag) comoving companion
at $2 \farcs 69$ west of \name{} at a projected separation of 94.1 au
(i.e. TIC 641991121, Gaia DR3 223158499176634112).
This source is clearly resolved by Gaia such that it cannot be
responsible for the excess noise in \name{'s} astrometric solution.
The companion does not appear in the 2MASS Point Source Catalog \citep{cutri03}.
Consequently, the 2MASS
blend and contamination/confusion flags for \name{} (\texttt{bb\_flg},
\texttt{cc\_flg}) indicate that its photometry was fit by a single source as it
was assumed to be uncontaminated. Similar issues of uncorrected
contamination persist for \name{} in all but the Gaia passbands. 
For inferring stellar parameters from empirical relations, we correct
\name{'s} $V$-band and 2MASS photometry using each source's Gaia photometry
and computing their magnitude differences in $VJHK_S$ using appropriate Gaia
color relations \citep{evans18}. We derive $\Delta$mag correction factors of
0.021, 0.080, 0.093, 0.099 in the $VJHK_S$-bands, respectively.

\begin{figure}
  \centering
  \includegraphics[width=\hsize]{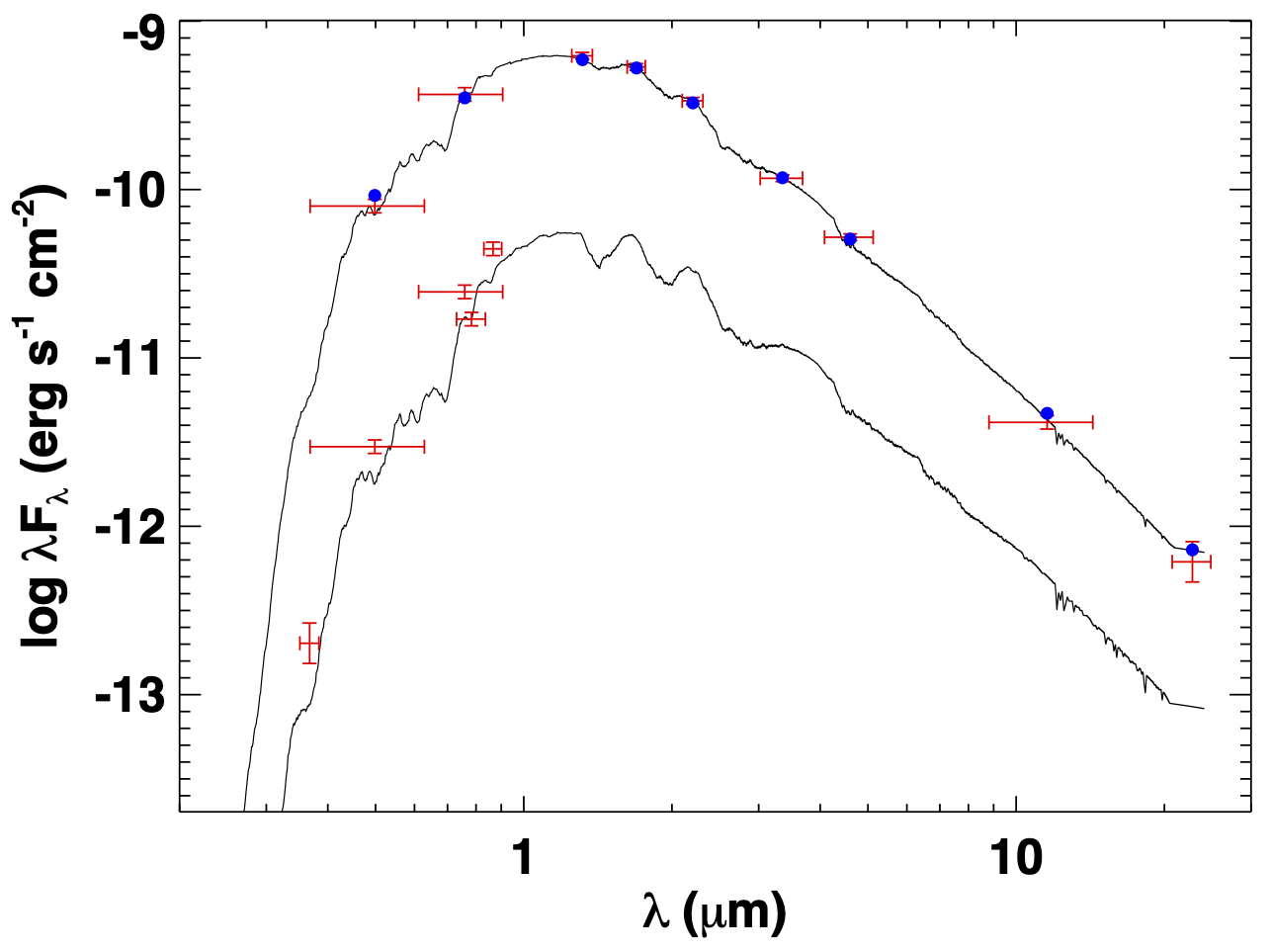}
  \caption{The spectral energy distributions of the target star \name{} and
    its faint companion. The black curves depict the stellar atmosphere models
    for each star with effective temperatures
    of 3500 K and 3025 K, respectively. The red markers depict the photometric
    measurements and their uncertainties. The horizontal errorbars depict the
    effective width of each passband. The blue markers depict the model
    flux in each passband for \name{.}}
  \label{fig:sed}
\end{figure}

The refined 2MASS photometry for \name{} has critical consequences for the
derivation of its global stellar properties from empirical relations. Using the
M dwarf $K_S$-band mass-luminosity relation from \cite{benedict16}, we find that
$M_s=0.502 \pm 0.014$ \Msun{.} This value is $1.3\sigma$ discrepant from the
result obtained without correcting the $K_S$-band magnitude. Similarly, we
measure a stellar radius of $R_s=0.450\pm 0.013$ \Rsun{} using the M dwarf
radius-luminosity relation from \cite{mann15}. Together these yield \logg{}
$=4.833\pm 0.028$. We derive the stellar effective temperature of \teff{}
$=3550 \pm 69$ K using the uncontaminated Gaia photometry and the
\teff{-}($G_{\text{BP}}-G_{\text{RP}}$) relation from \cite{mann15}.
We also estimate the stellar metallicity using the
empirical $(V-K_S)$-$M_{K_S}$-[Fe/H] relation for M dwarfs from
\cite{johnson09}. We find a somewhat metal-rich value of [Fe/H]
$=0.23^{+0.07}_{-0.08}$ dex, consistent with suggested correlations
for low mass stars between metallicity and the presence of small planets
\citep[e.g.][]{johnson09,schlaufman11}.

The companion star is in version 8 of the TESS Input Catalog
\citep[TIC;][]{stassun19}, with its TESS magnitude ($T=14.37$ mag) estimated
solely from Gaia photometry. We analyzed the spectral energy distributions
(SEDs)
of both stars to refine the dilution of \name{} in the TESS-band. Due to the
flux contamination, we performed a two-component fit following the procedures
outlined in \cite{stassun16,stassun17b,stassun18a}. For \name{,} we use the
$JHK_S$ magnitudes from 2MASS, W1--W4 from WISE, and Gaia
$G G_{\rm BP} G_{\rm RP}$ magnitudes. For the companion we use the $ui$-bands
from SDSS, the $y$-band from Pan-STARRS, and Gaia $G G_{\rm BP} G_{\rm RP}$
magnitudes (see Figure~\ref{fig:sed}). We fit for \teff{} and [Fe/H] in each SED
using a NextGen stellar atmosphere model \citep{hauschildt99} with zero
extinction ($A_V=0$). After correcting \name{'s} SED for the flux of the
companion, we measure \teff{} $=3500\pm 85$ K and [Fe/H] $=0.0\pm 0.5$ dex,
both of which are consistent with the values derived from empirical
relations. Similarly for the companion star, we measure
$T_{\text{eff,comp}}=3025\pm 100$ K and [Fe/H]$_{\rm comp}=0.0\pm 0.5$ dex.
Integrating the SED at a distance of 35.274 pc yields a bolometric
flux at Earth of $F_{\rm bol}=7.05\pm 0.27 \times 10^{-10}$ ergs s$^{-1}$
cm$^{-2}$, which corresponds to $R_s=0.452\pm 0.023$ \Rsun{} and again is
consistent with the value derived from the empirical
radius-luminosity relation. Given the total fluxes from our SED analysis, we
recover a dilution factor of $F_{1634} / (F_{1634}+F_{\text{companion}})= 0.946$,
which is consistent with the original value of 0.943 used by the NASA Ames
Science Processing Operations Center when producing the TESS light curve (see
Section~\ref{sect:tess}). Our derived dilution factor neglects the
two remaining sources that sit within the TESS aperture due to their
negligible flux contributions (see Figure~\ref{fig:tpf}).

The photometric stellar rotation period is presently unknown (see
Section~\ref{sect:asas}). We establish a prior on \prot{} using the
empirical M dwarf rotation-activity relation from \cite{astudillodefru17b}.  
From our HARPS-N spectra presented in Section~\ref{sect:harpsn}, we measure
\logrhk{} $=-5.39\pm 0.19$, which places \name{} within the unsaturated
regime of magnetic activity \citep[e.g.][]{reiners09b}. Using the
rotation-activity relation for inactive M dwarfs, we estimate
\prot{} $=77^{+26}_{-20}$ days.
Such a long rotation period
would place \name{} in the long-period tail of the \prot{} distribution among M
dwarfs with masses between $0.4-0.6$ \Msun{} \citep[$10-70$ days;][]{newton17}.

\section{Observations} \label{sect:observations}
\subsection{TESS Photometry} \label{sect:tess}
\name{} was observed by TESS for 24.38 days from UT 2019 November 3-27 in
Sector 18. The observations were taken with CCD 4 on camera 1.
\name{} is not slated for further observations with TESS\footnote{Based on the \href{https://heasarc.gsfc.nasa.gov/cgi-bin/tess/webtess/wtv.py}{TESS Web Viewing Tool}.}.
\name{} is listed in v8 of the TESS Input Catalog, the Candidate Target List
(CTL), and as a target in the Guest Investigator program
G022198\footnote{``\textit{Probing the Landscape of Cool Dwarf Planet Occurrence}''. PI: Dressing.} such that it was observed with 2-minute cadence. 
A total of 20 transits were observed with three
transit events being missed during the data transfer event near perigee
passage.

\begin{figure}
  \centering
  \includegraphics[width=\hsize]{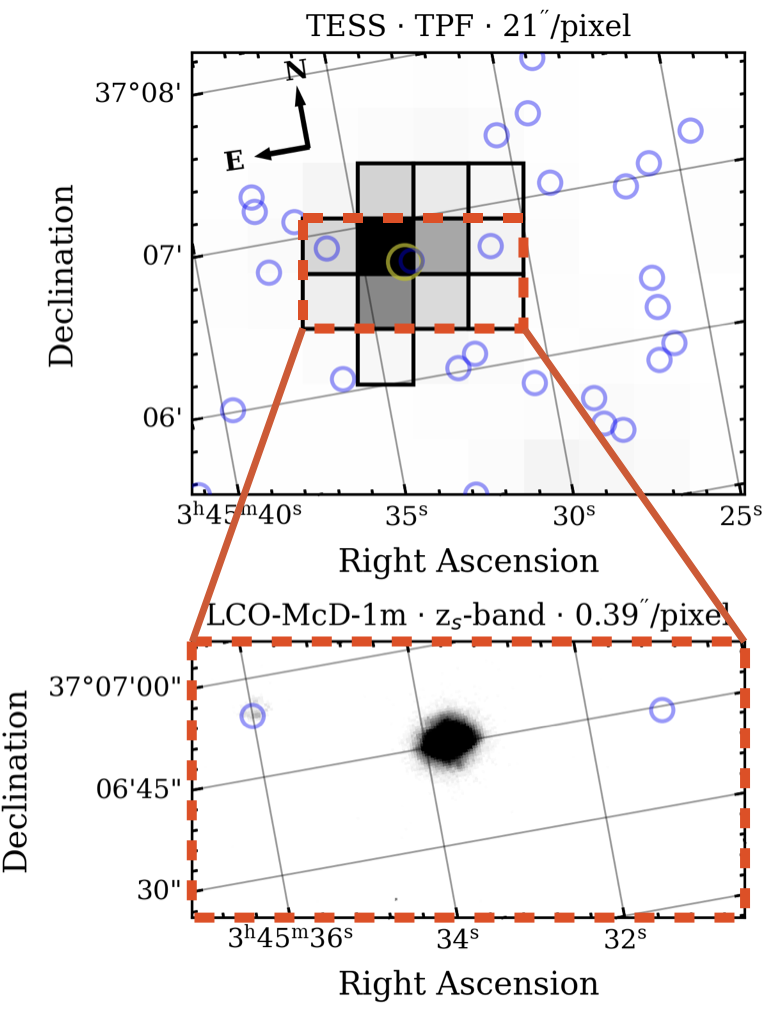}
  \caption{Images of the field surrounding \name{.} Upper panel: a sample TESS
    target pixel file image of \name{} with a pixel scale of $21\farcs$
    pixel$^{-1}$. The yellow circle highlights \name{} while the blue markers
    highlight its nearby stellar companion and other neighboring sources from
    Gaia
    EDR3. The pixels outlined in black demarcate the TESS photometric aperture
    used to produce the \texttt{PDCSAP} light curve of \name{.} Lower panel:
    a zoom in on the highlighted red region taken with the LCOGT 1m telescope
    at McDonald Observatory with a pixel scale of $0.39\farcs$. The small
    angular separation between \name{} and its companion prevent the source from
    being spatially resolved in our seeing-limited images.}
  \label{fig:tpf}
\end{figure}

A sample image from the TESS target pixel files (TPFs) is shown in the upper
panel of Figure~\ref{fig:tpf} overlaid by a subset of the 78 Gaia sources within
$2 \farcm 5$. All image data were processed by the NASA Ames Science Processing
Operations Center \citep[SPOC;][]{jenkins16}, who then proceeded to produce the
Presearch Data Conditioning Simple Aperture Photometry
\citep[\texttt{PDCSAP};][]{smith12,stumpe12,stumpe14} light curve using the
twelve-pixel photometric aperture overlaid in Figure~\ref{fig:tpf}. The aperture
clearly contains contributions from \name{,} its nearby stellar companion, and
at least two faint background sources from Gaia. \name{} dominates the
flux within the aperture and contributes 0.943 of the flux to the
\texttt{PDCSAP} light curve on average.

\begin{figure*}
  \centering
  \includegraphics[width=\hsize]{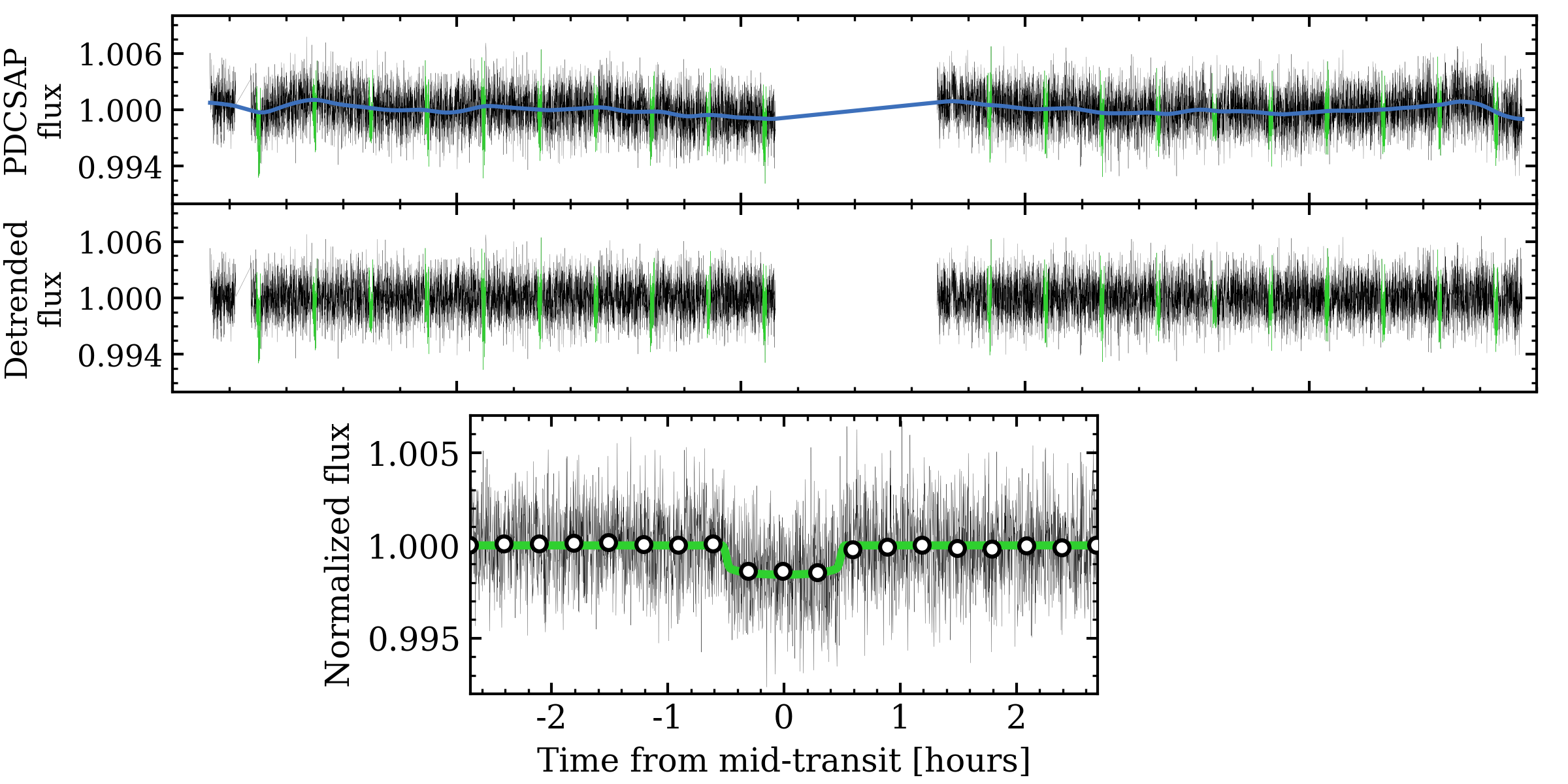}
  \caption{TESS \texttt{PDCSAP} light curve of \name{} from Sector 18. Top
    panel: the dilution and background-corrected \texttt{PDCSAP} light curve
    overlaid with the mean GP model of residual correlated noise (blue curve).
    In-transit measurements are highlighted in green. Middle panel: the
    \texttt{PDCSAP} light curve detrended by the mean GP model. Bottom panel:
    the phase-folded transit light curve of TOI-1634 b. The maximum
    a-posteriori transit model is overlaid in green while the white markers
    depict the binned light curve.}
  \label{fig:tess}
\end{figure*}

Late in the primary mission, the SPOC identified a bias in the
background sky correction that shifts the \texttt{PDCSAP} light curve to lower
flux values. Following the instructions outlined in the Sector 27 release
notes\footnote{\url{https://archive.stsci.edu/missions/tess/doc/tess_drn/tess_sector_27_drn38_v02.pdf}. The sky background algorithm was updated to mitigate the
  background bias starting with Sector 27.},
we correct this effect by determining the background bias
$bg_{\text{bias}}=9.35$ e$^{-}$/s/pixel from the difference between the
background-corrected pixel fluxes and zero. We then correct the \texttt{PDCSAP}
flux according to

\begin{equation}
  f'_{\text{PDCSAP}} = f_{\text{PDCSAP}} + bg_{\text{bias}}\: N_{\text{pix}}\:
  \frac{\text{CROWDSAP}}{\text{FLFRCSAP}},
\end{equation}

\noindent where $N_{\text{pix}}=12$ is the number of pixels in the
optimal aperture and CROWDSAP/FLFRCSAP $=1.14$ is the ratio of the crowding
metric to the flux fraction correction, which are provided in SPOC light
curve fits headers. This correction adjusts the baseline flux, and hence
decreases the inferred transit depth, by 2.2\%.

The dilution and background-corrected \texttt{PDCSAP} light curve for \name{}
is shown in the
upper panel of Figure~\ref{fig:tess} with the 20 transits of TOI-1634.01
highlighted in green. Note that no obvious signature of stellar rotation is
apparent in the light curve.
It is on these data that the SPOC conducted its transit search using the
Transiting Planet Search Pipeline Module \citep[TPS;][]{jenkins02,jenkins10}.
After passing a set of internal data validation tests \citep{twicken18,li19},
the TPS returned the new transiting planet candidate TOI-1634.01 with an orbital
period of 0.989 days and a transit depth of $1.52\pm 0.13$ ppt. Using
the stellar radius from Table~\ref{tab:star}, this initial transit depth
corresponds to a planet radius of $1.90\pm 0.10$ \Rearth{.} The public release
of the candidate TOI-1634.01 in December 2019 prompted our follow-up
observations described in the subsequent
Sections~\ref{sect:tres}-\ref{sect:harpsn}.

\subsection{Archival photometric monitoring} \label{sect:asas}
Recall that the TESS light curve does not show any signs of rotation
(Figure~\ref{fig:tess}). This is
consistent with \name{} being relatively inactive given its low
value of \logrhk{} $=-5.39\pm 0.19$ and the expectation of a long
rotation period \prot{}$=77^{+26}_{-20}$ days. Furthermore, we cannot hope to
obtain a precise \prot{} measurement with just one TESS sector if indeed
\prot{} is as long as we expect \citep{lu20}.

\begin{figure*}
  \centering
  \includegraphics[width=\hsize]{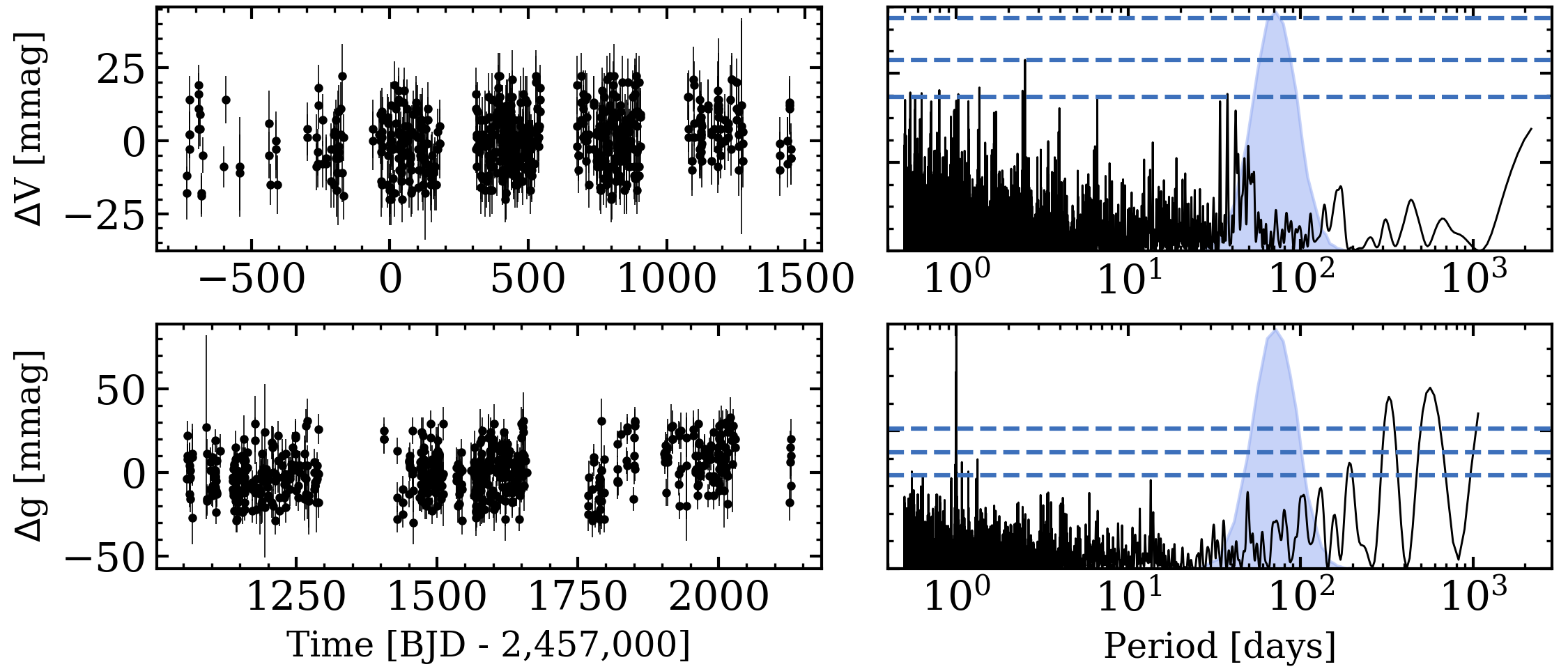}
  \caption{Photometric monitoring of \name{} with ASAS-SN in the $V$-band
    (upper row) and $g$-band (lower row). Left column: differential light
    curves. Right column: the GLS periodograms of each light curve. The blue
    histogram depicts the expected stellar rotation period based on the star's
    \logrhk{} and the rotation-activity relation from \cite{astudillodefru17b}.
    The horizontal dashed lines depict FAPs of 0.1\%, 1\%, and 10\%.
    No coherent periodic signal is detected.}
  \label{fig:asas}
\end{figure*}

We attempt to recover \prot{} by investigating the long-baseline
archival photometric monitoring from the ASAS-SN survey
\citep{jayasinghe19}. The ASAS-SN survey monitored \name{} from November 2012
to October 2020 in the $V$ and $g$-bands. Figure~\ref{fig:asas} shows the
light curves and their generalized Lomb-Scargle periodograms
\citep[GLS;][]{zechmeister09}. We compute the false alarm probability (FAP) for
each GLS periodogram via bootstrapping with replacement. We inspected the
periodogram of each light curve and found no coherent periodic signal that is
present in both light curves. Most notably, there is no persistent
signal over the domain between $50-105$ days where we expect to measure
\prot{} for \name{} based on its \logrhk{} value.

\subsection{Reconnaissance spectroscopy with TRES} \label{sect:tres}
Through the TESS Follow-up Observing Program (TFOP), we began to pursue the
confirmation of the planet candidate TOI-1634.01 by obtaining a pair of
reconnaissance spectra. We observed \name{} on UT 2020 February 2 and
2020 September 5 using the
Tillinghast Reflector \'{E}chelle Spectrograph (TRES). TRES is a fiber-fed
optical \'{e}chelle spectrograph (310-910 nm) with a resolution of $R=44,000$
and is mounted on the 1.5 m Tillinghast Reflector telescope at the Fred Lawrence
Whipple Observatory on Mount Hopkins, Arizona. The exposure time was set to
3000 s. We reduced and extracted the spectra using the standard
procedure \citep{buchhave10} before cross-correlating the spectra with a
custom spectral template of Barnard's star that was rotationally broadened over
a range of \vsini{} values\citep{winters18}.
We selected the \'echelle aperture 41 between $7065-7165 \AA$ for RV extraction
as it contains the information-rich TiO bands. We estimate the corresponding RV
precision at each epoch to be 65 and 38 \mps{.}

We find \name{} to be single-lined with no significant rotational broadening
(\vsini{}$<3.4$ \mps{)}, and with the $H\alpha$ feature in absorption
(Figure~\ref{fig:sg2}).
Our two TRES observations were also scheduled at opposing quadrature phases and
revealed no large RV variation beyond the level of our RV uncertainties.
These data confirm that \name{} is a chromospherically-inactive and
slowly-rotating star. These data also
likely rule out the possibility of a spectroscopic binary such
that TOI-1634.01 continues to be a viable planet candidate and we can
proceed with further attempts at planet confirmation.

\subsection{Seeing-limited photometry} \label{sect:sg1}
TESS pixels are large ($21''$), which results in blending of the \name{} light
curve with nearby sources. We therefore obtained seeing-limited photometry
to confirm the transit on-target and to spatially resolve the
light curves of nearby sources to rule out the nearby eclipsing binaries (NEBs)
as the source of the TESS transit events. We obtained a total of 16 light curves
of seven distinct transit events with a variety of observing facilities.
Table~\ref{tab:sg1} summarizes the observations with the individual facilities
described in the following sections. The light curves are shown in
Figure~\ref{fig:sg1}.

\begin{deluxetable*}{cccccc}
\tabletypesize{\small}
\tablecaption{Summary of seeing-limited photometric follow-up of \name{.}\label{tab:sg1}}
\tablewidth{0pt}
\tablehead{Obs. Date & Filter & Telescope & PSF FWHM & Photometric & Photometric  \\
  $[$YYYY-MM-DD$]$ && Aperture $[$m$]$ & $['']$ & Aperture $['']$ & Precision $[$ppt$]$\tablenotemark{a} }
\startdata
\multicolumn{6}{c}{\emph{LCO McDonald}} \\
2020-09-30 & $z_s$ & 1.0 & 4.2 & 2.5 & 0.5 \\
\hline
\multicolumn{6}{c}{\emph{MuSCAT2}} \\
2020-02-07 & $g,r,i,z_s$ & 1.52 & 1.9,1.8,1.8,1.7 & 4.0 & 2.6,0.9,1.0,0.8  \\ 
2020-02-10 & $g,r,i,z_s$ & 1.52 & 1.9,1.5,1.7,1.6 & 4.3 & 2.0,1.2,1.2,0.9  \\ 
2020-02-11 & $g,r,i,z_s$ & 1.52 & 1.8,1.5,1.6,1.2 & 4.3 & 1.6,1.1,0.8,0.8  \\ 
\hline 
\multicolumn{6}{c}{\emph{OAA}} \\
2020-02-13 & $I_c$ & 0.40 & 5.5 & 10.0 & 1.3 \\
2020-02-21 & $I_c$ & 0.40 & 7.6 & 10.0 & 1.4 \\
\hline
\multicolumn{6}{c}{\emph{RCO}} \\
2020-02-20 & $i'$ & 0.40 & 4.8 & 8.0 & 1.4 \\ 
\enddata
\tablenotetext{a}{Photometric precision is calculated as the rms of the detrended light curve in approximately 5-minute bins.}
\end{deluxetable*}

In summary, we successfully confirm the transit time of TOI-1634.01 and are able
to rule out 38 of 39 sources within $2\farcm 5$ as NEBs. However, the comoving
companion to \name{} at $2 \farcs 69$ is unresolved in all of our
observations (see lower panel of Figure~\ref{fig:tpf}). Even in our highest
quality ground-based light curves, at most 50\% of the companion's flux can
be excluded from the photometric aperture. 
As such, these data cannot uniquely identify \name{} as the host
of the TESS transit events, although they do limit the possibilities to either
\name{} or its companion.

\subsubsection{LCOGT}
We observed a full transit of TOI-1634.01 on UT 2020 September 30 in Pan-STARRS
$z_s$ band from the Las Cumbres Observatory Global Telescope
\citep[LCOGT;][]{brown13} 1 m network node at McDonald Observatory. We used the
\texttt{TESS Transit Finder}, which is a customized version of the
\texttt{Tapir} software package \citep{jensen13}, to schedule our transit
observations. The $4096\times 4096$ LCOGT SINISTRO cameras have an image
scale of $0 \farcs 39$ per pixel, resulting in a
$26 \arcmin \times 26 \arcmin$ field of view. The images were calibrated
by the standard LCOGT \texttt{BANZAI} pipeline \citep{mccully18}, and
photometric data were extracted with \texttt{AstroImageJ} \citep{collins17}. The
TOI-1634.01 observation used 40 second exposures and a photometric aperture
radius of $2\farcs 5$ to extract the differential photometry.

\subsubsection{MuSCAT2}
MuSCAT2 \citep{narita19} is a multi-color camera that is able to obtain
simultaneous observations in four bands: \textit{Sloan-g}, \textit{Sloan-r},
\textit{Sloan-i}, and \textit{Sloan-$z_s$}. The instrument is mounted on the
1.52m Telescopio Carlos S\'{a}nchez (TCS) at Teide Observatory, Tenerife, Spain.
The field of view of MuSCAT2 is $7.4 \arcmin \times 7.4 \arcmin$ with a pixel
scale of $0 \farcs 44$ per pixel. All the cameras have a short read out time
between 1-4 seconds, which makes MuSCAT2 an ideal instrument for transit
follow-up and time-series observations in general. We observed three primary
transits of TOI-1634b in all four bands on the nights of UT 2020 February
7, 10, and 11. For each night, we set the exposure times to avoid the
saturation of the target star. We reduced the data using standard procedures:
the photometry and transit model fit (including systematic effects) was done by
the MuSCAT2 pipeline \citep{parviainen19,parviainen20}.

\subsubsection{OAA}
We observed two full transits of TOI-1634.01 on UT 2020 February 13 and 21 using
the main 0.4 m instrument ensemble at Observatori Astron\`omic Albany\`a (OAA)
with stable observation conditions in the valley. We performed differential
photometry in a $36 \arcmin \times 36 \arcmin$ star field centered on \name{}
using the $I_c$ filter with $10 \farcs 0$ photometric apertures (in $7 \farcs 4$
FWHM conditions) using the \texttt{AstroImageJ} pipeline. The sequences
consisted of 88 and 148 frames of 120 s and 100 s exposure times, respectively.
A small number of outlying
points during transit due to instrumental inconveniences ($>$10$\sigma$) were
removed before the transit fit. No significant NEB signals were detected 
within $2 \farcm 5$ of the target after performing a thorough NEB check with
different apertures from $4 \farcs 5 - 10 \farcs 0$.

\subsubsection{RCO}
A full transit observation of TOI-1634.01 was obtained on UT 2020 February 20
using the RCO 40 cm telescope located at the Grand-Pra Observatory, Switzerland.
We observed a full transit in the \textit{Sloan i'} passband with an exposure
time of 90 seconds. We produced the light curve of TOI-1634.01 using the
\texttt{AstroImageJ} pipeline with $8\farcs 0$ apertures and by detrending
against airmass and FWHM. We confirmed that no NEB signals appeared at the
expected time within $2 \farcm 5$ of \name{.}

\subsection{High-resolution imaging} \label{sect:sg3}
The smallest PSF of our seeing-limited photometric observations has a full
width at half maximum (FWHM) of $1\farcs 2$.
Thus, with seeing-limited photometry alone we are insensitive to sources more
closely separated from \name{} than approximately this limit. To check for
blended sources within $1\farcs 0$, we obtained four sets of
high-resolution imaging sequences (Figure~\ref{fig:sg3}), which
are described in the following subsections.
Other than the known stellar companion at $2\farcs 69$ separation, we do not
find evidence for any additional contaminating sources down to $0\farcs 2$ given
the sensitivity of our observations and thus do not find any supporting
evidence for a massive long-period companion that may have been able to
account for \name{'s} excess astrometric noise in Gaia EDR3.
As such, TOI-1634.01 remains a viable planet candidate.

\subsubsection{$'$Alopeke}
We obtained speckle interferometric images of \name{} on UT 2020 February 16
using the $'$Alopeke
instrument\footnote{https://www.gemini.edu/instrumentation/alopeke-zorro}
mounted on the 8 m Gemini North telescope on the summit of Maunakea in Hawai'i.
$'$Alopeke simultaneously collects diffraction-limited images at 562 and 832 nm.
Our data set consists of 7 minutes of total integration time taken as sets of
$1000 \times 0.06$ s images. Following \cite{howell11}, we combined all images,
subjected them to Fourier analysis, and produced reconstructed images from
which the $5\sigma$ contrast curves are derived in each passband.
Figure~\ref{fig:sg3} presents the two contrast curves as well as the 832 nm
reconstructed image. Our measurements reveal \name{} to be a single star down
to $\Delta$mag $5-7$, eliminating all main sequence stellar companions
earlier than M6 within the spatial limits of 0.6-1.0 au at the inner working
angle, and out to 42 au at $1\farcs 2$.

\subsubsection{ShARCS}
We observed \name{} on UT 2020 December 1 using the ShARCS camera on the Shane
3m telescope at Lick Observatory. Our observations were taken using the Shane
adaptive optics (AO) system in natural guide star mode. We collected our
observations using a 4-point dither pattern with a separation of $4\farcs 0$
between
each dither position. We obtained a pair of sequences; in the $J$ and $K_S$
bands with exposure times of 7.5 s and 15 s, respectively. See
\cite{savel20} for a detailed description of the observing strategy and
reduction procedure. Our AO images and contrast curves for each imaging sequence
are shown in Figure~\ref{fig:sg3}. We detect the known companion
but find no other nearby companions within $1\farcs 0$ down to $\Delta J=3.6$
mag and $\Delta K_S=4$ mag.

\subsection{Precise radial velocity measurements} \label{sect:harpsn}
We obtained 32 spectra of \name{} using the HARPS-N spectrograph located at the
3.6 m Telescopio Nazionale Galileo (TNG) on La Palma, Canary Islands.
HARPS-N is a high resolution ($R=115,000$) optical \'{e}chelle spectrograph
whose long-term pressure and temperature stability enable it to reach
sub-meter-per-second stability \citep{consentino12}. The exposure
time was fixed to 1800 s. We follow the standard procedure for M dwarf
observations with HARPS-N and focus solely on the \'{e}chelle orders redward
of aperture 18 \citep[i.e. 440-687 nm;][]{anglada12}.
The median total S/N of our spectra is 107.

We obtained our observations over a 210-day span between UT 2020 August 7 and
2021 March 4 
as part of the HARPS-N collaboration Guaranteed Time Observations. Due to the
proximity of TOI-1634.01's orbital period to one day
($P=0.989$ days), we were unable to obtain uniform sampling of the planet's
orbital phase. Fortunately, the combination of the planet's ephemeris and the
longitude of the TNG observatory resulted in preferential sampling
of the planet's orbit near its quadrature phases (i.e. $\phi \sim \pm 0.25$).
The information content of our time series with respect to the RV semiamplitude
is much richer than if only
orbital phases close to 0 and 0.5 could be sampled. However, although this
restricted sampling had only a small effect on the inference of the planet's
RV semiamplitude from preliminary analyses, we found that the constraints on
the orbital eccentricity were very weak when left unconstrained. Note that
given the planet's ultra-short period (USP),
it is reasonable to expect a circularized orbit with little
to no eccentricity (see Section~\ref{sect:mr}).
To remedy the lack of observational constraints on orbital eccentricity, the
six most recent RV measurements were
intentionally scheduled to fill in the gaps in our orbital phase sampling in an
effort to distinguish between circular and eccentric orbital solutions.

We extracted the RVs via template-matching using the \texttt{TERRA} pipeline
\citep{anglada12}. Template-matching is a commonly used tool for the RV
extraction from M dwarf spectra as it is known to achieve improved RV precision
over the more traditional cross-correlation function techniques
\citep[e.g.][]{astudillodefru15}. \texttt{TERRA} works by constructing a
master template spectrum by coadding all of the individual spectra after
shifting each spectrum to the barycentric frame. The barycentric corrections are
retrieved from the HARPS-N Data Reduction Software
\citep[DRS;][]{lovis07}. We ignore spectral regions in which the telluric
absorption exceeds 1\%. The RV of each spectrum is then calculated
via least-squares matching of the spectrum to the master template in velocity
space. Due to the poor S/N of the bluest orders, we only focus on \'{e}chelle
orders redward of aperture 18. We obtain a median RV uncertainty of 1.73 \mps{.}

\begin{figure}
  \centering
  \includegraphics[width=\hsize]{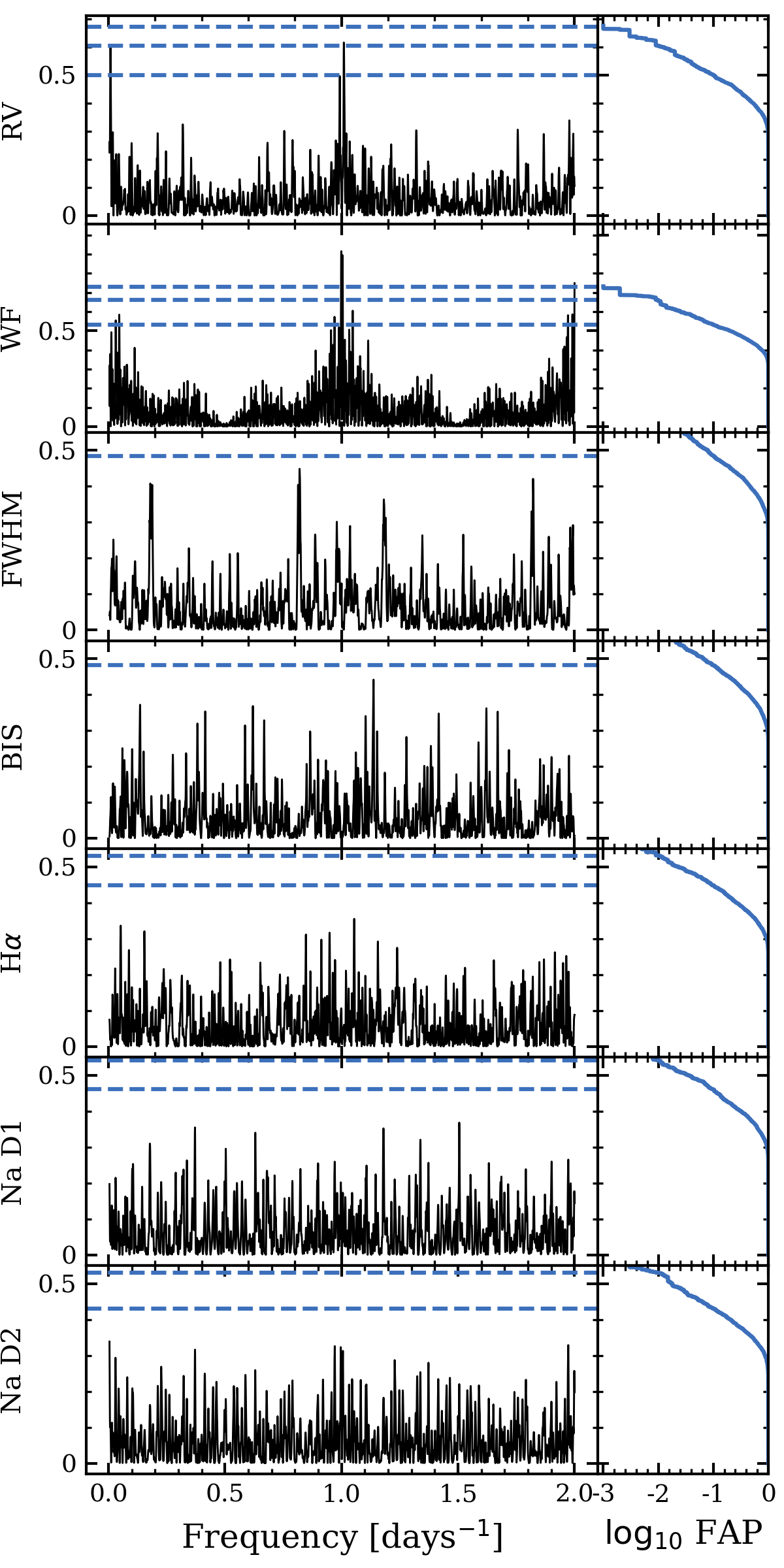}
  \caption{GLS periodograms of the HARPS-N RVs, window function (WF), and
    spectroscopic indicators of \name{.} Left column: the GLS periodogram of
    the time series labeled on the y-axis. The horizontal dashed lines report
    the FAP levels of 0.1\%, 1\%, and 10\%. Right column: the FAP as a function
    of normalized power.}
  \label{fig:gls}
\end{figure}

\begin{figure}
  \centering
  \includegraphics[width=\hsize]{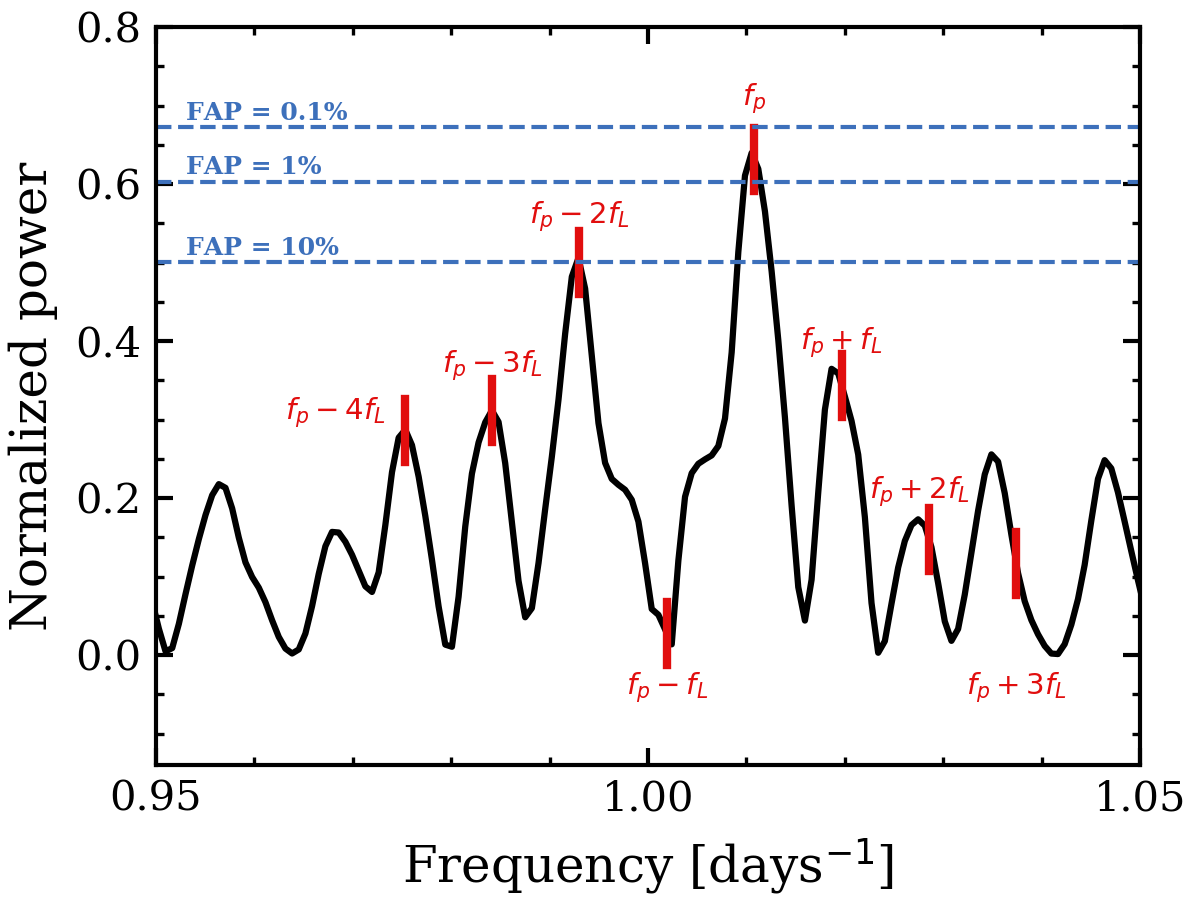}
  \caption{The GLS periodogram of the \name{} RVs in the vicinity of the
    planet candidate's orbital frequency $f_p = 1/0.98932$ days$^{-1}$. The
    forest of peaks can be well-explained as aliasing by the long period
    frequency $f_L=1/113$ days$^{-1}$ seen in the RVs. The horizontal dashed
    lines report the FAP levels of 0.1\%, 1\%, and 10\%.}
  \label{fig:glszoom}
\end{figure}

Figure~\ref{fig:gls} shows the GLS periodograms of the RVs, the window function,
and the following activity indicators produced by the DRS: CCF FWHM, CCF BIS,
H$\alpha$ ($6563 \AA$), and both sodium doublet features Na D1 ($5890 \AA$) and
Na D2 ($5896 \AA$). We do not observe any significant periodic signals in any
of the activity indicators, thus we do not recover the stellar rotation period
from these spectroscopic indicators. The only significant (FAP $< 1$\%)
persistent signal that emerges in multiple time series is close to 1-day, which
we expect in the RVs due to the transiting planet candidate with $P=0.989$ days
(also shown zoomed-in in Figure~\ref{fig:glszoom}). The 1-day signal is
also apparent in the window function due to effect of the one day alias: a
phenomenon that often
inhibits the detection of periodic RV signals close to one day \citep{dawson10}
but is not a major issue in our analysis due to the strong prior
on the planet's orbital period from the transit data. The time series depicted
in Figure~\ref{fig:gls} are provided in Table~\ref{tab:rv}.

\begin{deluxetable*}{cccccccc}
\tablecaption{HARPS-N time series of \name{}\label{tab:rv}}
\tablewidth{0pt}
\tablehead{Time & RV & $\sigma_{\text{RV}}$ & FWHM & BIS & H$\alpha$ & Na D1 & Na D2 \\
$[$BJD - 2,457,000$]$ & $[\text{m s}^{-1}]$ & $[\text{m s}^{-1}]$ & $[\text{km s}^{-1}]$ & $[\text{km s}^{-1}]$ & & &}
\startdata
2068.720217 & -0.246 & 1.753 & 3.064 & -12.248 & 0.911 & 0.486 & 0.613 \\
2069.716163 & -6.512 & 2.427 & 2.974 & -6.289 & 0.892 & 0.472 & 0.627 \\
2070.717423 & -6.615 & 1.572 & 2.989 & 4.337 & 0.879 & 0.441 & 0.585 \\
\enddata
\tablecomments{For conciseness, only a subset of three rows are depicted here to illustrate the table's contents. The entirety of this table is provided in the arXiv source code and will ultimately be available as a machine readable table in the journal.}
\end{deluxetable*}

One remaining low FAP signal is seen solely in the RVs at a frequency of
$1/113=0.00885$ days$^{-1}$. The origin of the 113-day signal is
unlikely to be due to stellar rotation as the signal is not visible in any of
the activity indicators, and it would represent an uncharacteristically long
rotation period for an inactive M dwarf with the mass of \name{.}
The signal may also potentially be due to the long-period companion
that was posited based on the excess noise in the Gaia EDR3 astrometry
(Section~\ref{sect:star}). However, Figure~\ref{fig:glszoom} reveals
that the 113-day signal (i.e. $f_L = 1/113=0.00885$ days$^{-1}$) is an alias
as it can explain the forest of peaks aliasing the planet candidate at the
frequencies $f_p+nf_L$, where $f_p=1/0.98934=1.01077$ days$^{-1}$ is the
orbital frequency of TOI-1634.01 and $n$ takes on integer values. In
Section~\ref{sect:global} we will confirm that the 113-day signal is a spurious
aliased signal that disappears upon the removal of the signal at $f_p$.

\section{Transit Plus RV Analysis and Results} \label{sect:analysis}
We proceed with measuring the accessible planetary parameters following a
two-step process. We first model the TESS transit light curve alone to remove
any residual low-order systematics and to derive initial estimates of the
transit parameters (Section~\ref{sect:transit}). We then use those
initializations to produce a global transit plus RV
model from which we measure the physical and orbital properties of \name{} b
(Section~\ref{sect:global}).

\subsection{TESS Transit Analysis} \label{sect:transit}
Standard systematics detrending has already been applied to the TESS
\texttt{PDCSAP} photometry by the SPOC. However, some low-amplitude variability
is seen to persist which we attribute to residual systematics
(top panel of Figure~\ref{fig:tess}). Here we model the \texttt{PDCSAP} light
curve with a transiting planet model plus systematics model in the form of an
untrained Gaussian process (GP). The covariance of the GP is parameterized as a
stochastically-driven simple harmonic oscillator in Fourier space, which enables
efficient computations of the GP's marginalized likelihood when operating on
large datasets (i.e. when number of data points $\gg$ number of model
parameters). The spectral density of the covariance kernel is

\begin{equation}
  S(\omega) = \sqrt{\frac{2}{\pi}} \frac{S_0 \omega_0^4}{(\omega^2 - \omega_0^2)^2 + (\sqrt{2} \omega_0 \omega)^2},
\end{equation}

\noindent where $\omega_0$ is the frequency of the undamped oscillator and $S_0$
describes the spectral power at $\omega_0$. We also include an additive scalar
jitter term to account for any excess uncorrelated noise in the TESS photometry:
$s_{\text{TESS}}$. The simultaneous \cite{mandel02}
transit model has the following free
parameters: stellar mass $M_s$, stellar radius $R_s$, quadratic limb-darkening
coefficients ${u_{1,T}, u_{2,T}}$,
orbital period $P$, time of mid-transit $T_0$, planet
radius $r_p$, impact parameter $b$, eccentricity $e$, argument of periastron
$\omega_p$, and flux baseline $f_0$. We include samples of $M_s$ and $R_s$ as,
together with $P$, they uniquely constrain the scaled semimajor axis
$a/R_s$ and the stellar density, which in turn constrains permissible values
of $e$ and $\omega_p$ \citep{moorhead11,dawson12}. Our full model features 14
model parameters with the following parameterizations:
$\{\ln{\omega_0}, \ln{S_0 \omega_0^4}, \ln{s_{\text{TESS}}^2}, M_s, R_s, u_{1,T}, u_{2,T}, \ln{P}, T_0, \ln{r_p},$
$b, e, \omega_p, f_0 \}$. The respective priors are listed
in Table~\ref{tab:priors}.

\begin{deluxetable}{lc}
\tabletypesize{\small}
\tablecaption{\tess{} light curve and RV model parameter priors\label{tab:priors}}
\tablewidth{0pt}
\tablehead{Parameter & Fiducial Model Priors}
\startdata
\multicolumn{2}{c}{\emph{Stellar parameters}} \\
$M_s$ [\Msun{]} & $\mathcal{N}(0.502,0.014)$ \\
$R_s$ [\Rsun{]} & $\mathcal{N}(0.450,0.013)$ \\
\multicolumn{2}{c}{\emph{Light curve hyperparameters}} \\
$f_{0,T}$ & $\mathcal{U}(-\inf,\inf)$ \\
$\ln{\omega_0}$ [days$^{-1}$] & $\mathcal{N}(0,10)$  \\
$\ln{S_0 \omega_{0}^{4}}$ & $\mathcal{N}(\ln{\text{var}(f'_{\rm PDCSAP})},10)$ \\
$\ln{s_{\text{TESS}}^2}$ & $\mathcal{N}(\ln{\text{var}(f'_{\rm PDCSAP})},10)$\tablenotemark{a} \\
$u_{1,T}$ & $\mathcal{U}(0,1)$  \\
$u_{2,T}$ & $\mathcal{U}(0,1)$ \\
\multicolumn{2}{c}{\emph{RV parameters}} \\
$\ln{s}_{\rm RV}$ [\mps{]} & $\mathcal{U}(-5,5)$ \\
$\gamma_{\rm RV}$ [\mps{]} & $\mathcal{U}(-10,10)$  \\
\multicolumn{2}{c}{\emph{\name{} b parameters}} \\
$P$ [days] & $\mathcal{U}(-\inf,\inf)$ \\
$T_{0}$ [BJD-2,457,000] & $\mathcal{U}(-\inf,\inf)$  \\
$\ln{r_p}$ [\Rearth{]} & $\mathcal{N}(0.5\cdot \ln(Z) + \ln{R_s},1)$\tablenotemark{b} \\
$r_{p}/R_s$  & $\mathcal{U}(-\inf,\inf)$ \\
$b$ & $\mathcal{U}(0,1+r_p/R_s)$ \\
$\ln{K}$ [\mps{]} & $\mathcal{U}(-4,4)$ \\
$e$ & $\mathcal{B}(0.867,3.03)$\tablenotemark{c} \\
$\omega$ [rad] & $\mathcal{U}(-\pi,\pi)$\tablenotemark{c} \\
$\sqrt{e}\cos{\omega}$ & $\mathcal{U}(-1,1)$ \\
$\sqrt{e}\sin{\omega}$ & $\mathcal{U}(-1,1)$ \\
\enddata
\tablecomments{Gaussian distributions are denoted by $\mathcal{N}$ and are
  parameterized by mean and standard deviation values. Uniform distributions
  are denoted by $\mathcal{U}$ and bounded by the specified lower and upper
  limits. Beta distributions are denoted by $\mathcal{B}$ and are parameterized
  by the shape parameters $\alpha$ and $\beta$.}
\tablenotetext{a}{$f'_{\rm PDCSAP}$ is the flux time series representing the dilution and background-corrected \texttt{PDCSAP} light curve from TESS.}
\tablenotetext{b}{The transit depth of TOI-1634.01 reported by the SPOC: $Z=1520$ ppm.}
\tablenotetext{c}{For use in the TESS analysis only \citealt{kipping13}.}
\end{deluxetable}

We use \texttt{PyMC3} \citep{salvatier16} within the
\texttt{exoplanet} package \citep{foremanmackey19} to evaluate the model's
joint posterior via Markov Chain Monte Carlo (MCMC). Within
\texttt{exoplanet}, the separate software packages \texttt{celerite}
\citep{foremanmackey17} and \texttt{STARRY} \citep{luger19} are used to
calculate the GP and transit models, respectively. We run four simultaneous
chains with 4000 tuning steps to derive the model's joint posterior. We use
the maximum a-posteriori (MAP) point estimates of the GP hyperparameters to
construct the GP posterior (i.e. predictive) distribution whose mean function
we use to detrend the TESS photometry (middle panel of Figure~\ref{fig:tess}).
We then adopt the MAP transit model parameters to initialize the MCMC of our
global model in the next section.

\subsection{Global Modeling} \label{sect:global}
We proceed with constructing our global model, which jointly considers the
transit and RV datasets. The primary
purpose of our seeing-limited photometric observations (Section~\ref{sect:sg1})
were to rule out neighboring sources as the origin of the TESS transit
events (i.e. NEBs). This purpose has been successfully served so there is no
need to include all of those observations in our global model.
Instead, here we only include the most recent high S/N observation from
LCOGT in the $z_s$-band.
This choice provides the longest time baseline and thus provides the
strongest constraints on the planet's ephemeris.

Even inactive M dwarfs rotate and exhibit some level of magnetic activity.
However, our photometric and spectroscopic
analyses have indicated that \name{} shows no evidence
for coherent and temporally sustained signals from stellar activity. As such,
in our fiducial model, we do not attempt to model any temporal
correlations from stellar activity and simply model excess jitter with an
additive scalar
term $s_{\rm RV}$. Our fiducial transit plus RV model therefore features a total
of 16 parameters. Among these are the
same transit model parameters described in Section~\ref{sect:transit}, with the
exception of the GP hyperparameters as here we consider the detrended TESS
light curve. However, we modify the parameterization of $r_p$, $e$, and
$\omega_p$ as follows. The planet radius $r_p$ becomes the planet-to-star ratio
$(r_p/R_s)_i$, which has a unique index $i$ for each passband $\in [T,z_s]$.
Similarly, each passband has
a unique flux baseline $f_{0,i}$.
The $z_s$ limb-darkening coefficients were fixed to $u_{1,LCO}=0.17$ and
$u_{2,LCO}=0.42$ \citep{claret11}. To avoid the Lucy-Sweeney bias against $e=0$,
we elect to sample the parameters $h=\sqrt{e}\cos{\omega_p}$ and
$k=\sqrt{e}\sin{\omega_p}$ \citep{lucy71,eastman13}\footnote{Due to the
  high probability of \name{} b being tidally circularized,
  we also tested a model with a fixed circular orbit and found that the
  resulting RV semiamplitude and its measurement precision are effectively
  insensitive to the assumption of a circular orbit.}.
The RV component of our model then
consists of three additional parameters: the RV semiamplitude $K$, the 
velocity offset $\gamma_{\rm RV}$, and the aforementioned additive
scalar jitter $s_{\rm RV}$. Our complete set of model parameters is
$\{ M_s, R_s, f_{0,T}, f_{0,LCO}, u_{1,T}, u_{2,T}, P, T_0, b, (r_p/R_s)_T,$
$(r_p/R_s)_{LCO}, h, k, \ln{K}, \gamma_{\rm RV}, \ln{s_{\rm RV}} \}$.
Their respective priors are also listed in Table~\ref{tab:priors}.

Given that the Gaia EDR3 astrometric solution may be consistent with the
existence of a long-period companion, we also considered an RV model that
includes a linear trend term. We determine that the slope of the linear
trend is consistent with zero, thus indicating that our RV data are able to
strongly rule out a long-period companion out to approximately the baseline
of our observations (i.e. 210 days).

\begin{figure*}
  \centering
  \includegraphics[width=\hsize]{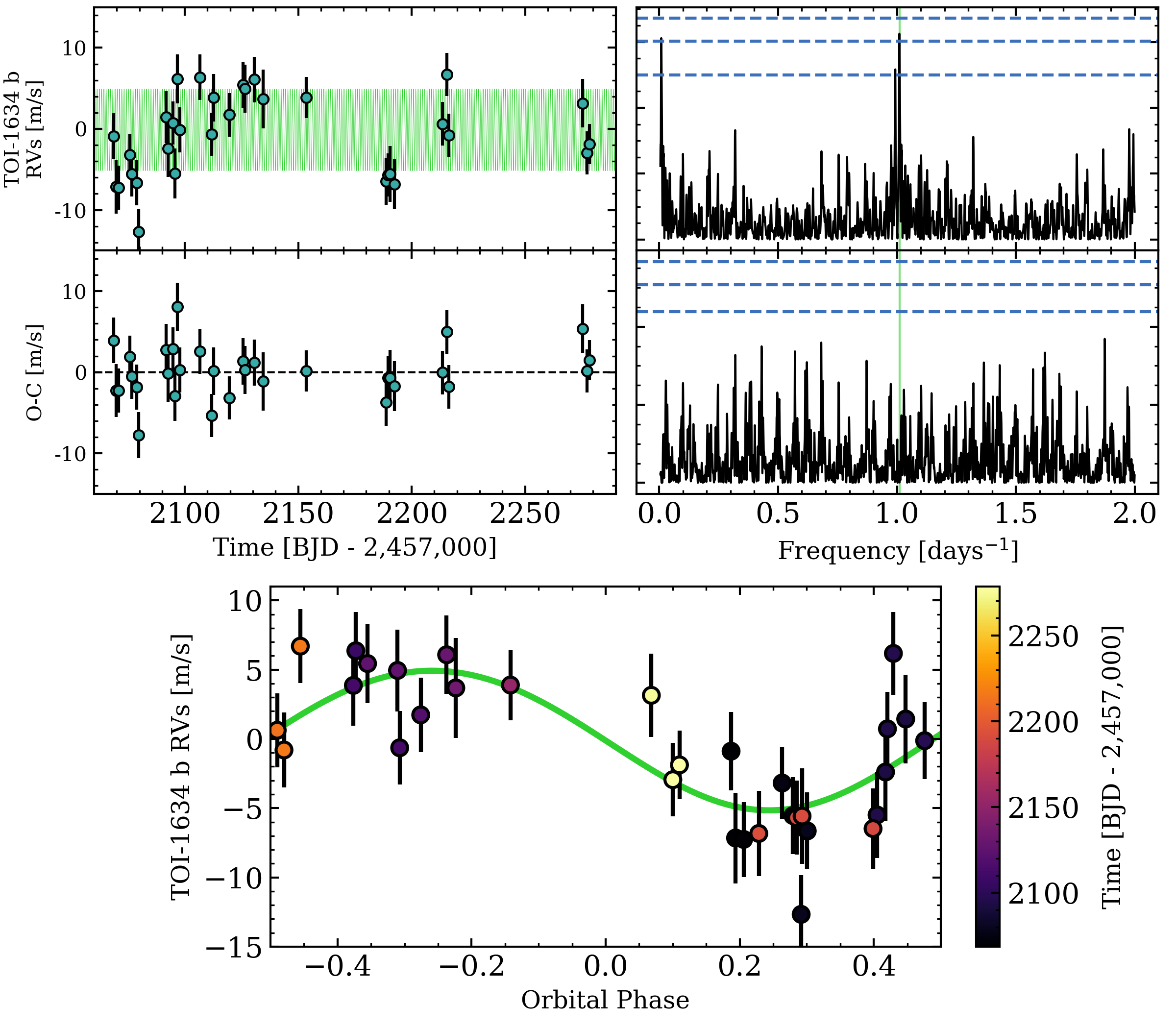}
  \caption{The \name{} RVs and model from our fiducial global analysis.
    Top row: the raw HARPS-N RVs overlaid with the best-fit Keplerian
    solution for \name{} b. The GLS periodogram of the RVs is shown on the left.
    The vertical green band highlights the orbital period of \name{} b. The
    horizontal dashed lines depict the 0.1\%, 1\%, and 10\% FAPs. Middle row:
    the RV residuals along with the corresponding GLS periodogram. Bottom panel:
    the planetary signal phase-folded to the orbital period of \name{} b. The
    marker colors indicate the individual observation times, which illustrates
    our effort to obtain more complete sampling of the orbital phase. The
    RV measurement uncertainties throughout include the contribution from the
    additive scalar RV parameter $s_{\rm RV}$.}
  \label{fig:results}
\end{figure*}

We fit the TESS, LCO, and HARPS-N RV data with our fiducial model and sample
the joint posterior using the
affine-invariant ensemble MCMC sampler \texttt{emcee} \citep{foremanmackey13}.
We initialize 200 walkers and evaluate the convergence of each walker's chain
by insisting that $\geq 10$ autocorrelation times are sampled.
MAP point estimates of the model parameters are derived from their respective
marginalized posteriors and are reported in Table~\ref{tab:results}
along with uncertainties derived from the $16^{\rm th}$ and $84^{\rm th}$
percentiles. The resulting transit model is shown in the lower panel of
Figure~\ref{fig:tess} while the RV results are shown in
Figure~\ref{fig:results}. The Keplerian RV signal from \name{} b is clearly
detected with a semiamplitude of $K=5.04^{+0.70}_{-0.72}$ \mps{} and on an orbit
that is consistent with circular (i.e. $e<0.16$ at 95\% confidence). The
113-day signal in the RVs disappears with the subtraction of the
planet model, which supports the notion that the signal was merely an alias
rather than physical.

Notably, we find the MAP scalar jitter to be comparable to the median
RV measurement uncertainty ($s_{\rm RV} = 2.2\pm 0.5$ \mps{).} This indicates
that there is a significant dispersion in the RVs that is unrelated to the
known planet and does not exhibit a coherent periodicity. We note that we
consider the possibilities of stellar activity and additional planets in
Sections~\ref{sect:act} and~\ref{sect:sens}. With the quadrature addition of
$s_{\rm RV}$ to the RV uncertainties, our RV residuals exhibit an rms of
3.10 \mps{} with $\chi^2=1.21$.

\startlongtable
\begin{deluxetable}{lc}
\tabletypesize{\footnotesize}
\tablecaption{Point estimates of the \name{} model parameters\label{tab:results}}
\tablewidth{0pt}
\tablehead{Parameter & Fiducial Model Values}
\startdata
\multicolumn{2}{c}{\emph{Transit parameters}} \\
Baseline flux, $f_{0,T}$ & $1.000035\pm 0.000020$  \\
Baseline flux, $f_{0,LCO}$ & $1.0015\pm 0.0026$  \\
$\ln{\omega_0}$ & $0.85\pm 0.10$ \\
$\ln{S_0 \omega_0^4}$ & $1.54^{+0.51}_{-0.56}$  \\
$\ln{s_{\text{TESS}}^2}$ & $0.0014\pm 0.006$ \\
\tess{} limb darkening  & $0.28^{+0.19}_{-0.15}$ \\
\hspace{12pt} coefficient, $u_1$ & \\
\tess{} limb darkening & $0.09^{+0.22}_{-0.20}$ \\
\hspace{12pt} coefficient, $u_2$ & \\
\multicolumn{2}{c}{\emph{RV parameters}} \\
Log Jitter, $\ln{s_{\rm RV}}$ & $0.81\pm 0.17$ \\
Velocity offset, $\gamma_{\rm RV}$ [\mps{]} & $0.32^{+0.56}_{-0.58}$ \\
\multicolumn{2}{c}{\emph{TOI-1634 b parameters}} \\
Orbital period, $P$ [days] & $0.989343\pm 0.000015$ \\
Time of mid-transit, & $1791.51473\pm 0.00061$ \\
\hspace{12pt} $T_0$ [BJD - 2,457,000] & \\
Transit duration $D$ [hrs] & $1.027\pm 0.028$ \\
Transit depth, $Z$ [ppt] & $1.323^{+0.095}_{-0.092}$ \\
Scaled semimajor axis, $a/R_s$ & $7.38\pm 0.20$ \\
Planet-to-star radius ratio, $r_p/R_s$ & $0.0364\pm 0.0013$ \\
Impact parameter, $b$ & $0.24\pm 0.13$ \\
Inclination, $i$ [deg] & $88.2\pm 1.1$ \\
Eccentricity, $e$ & $<0.16$\tablenotemark{a} \\
Planet radius, $r_p$ [\Rearth{]} & \rplanet{} \\
Log RV semiamplitude, $\ln{K}$ & $1.62^{+0.13}_{-0.15}$ \\
RV semiamplitude, $K$ [\mps{]} & $5.04^{+0.70}_{-0.72}$ \\
Planet mass, $m_p$ [\Mearth{]} & \mplanet{} \\
Bulk density, $\rho_p$ [g cm$^{-3}$] & $4.7^{+1.0}_{-0.9}$ \\
Surface gravity, $g_p$ [m s$^{-2}$] & $15.0^{+2.6}_{-2.5}$ \\
Escape velocity, $v_{\rm esc}$ [km s$^{-1}$] & $18.5^{+1.3}_{-1.4}$ \\
Semimajor axis, $a$ [au] & $0.01545\pm 0.00014$ \\
Insolation, $F$ [F$_{\oplus}$] & $121^{+12}_{-11}$ \\
Equilibrium dayside temperature, & $1307\pm 30$ \\
\hspace{12pt} $T_{\text{eq,day}}$ [K]\tablenotemark{b} & \\
Equilibrium temperature, $T_{\text{eq}}$ [K]\tablenotemark{c} & $924\pm 22$ \\
Envelope mass fraction, $X_{\rm env}$ [\%]\tablenotemark{d} & $0.30^{+0.19}_{-0.17}$ \\
\enddata
\tablenotetext{a}{95\% upper limit.}
\tablenotetext{b}{Assuming a tidally locked dayside and zero albedo.}
\tablenotetext{c}{Assuming uniform heat redistribution and zero albedo.}
\tablenotetext{d}{Assuming an Earth-like solid core with a 33\% iron core mass fraction (i.e. a 33\% iron inner core plus a 67\% silicate mantle).}
\end{deluxetable}

\subsection{Attempts at more sophisticated treatments of stellar activity} \label{sect:act}
We note that we did make additional attempts at more complete RV models that
included a treatment of evolving stellar activity. Our first attempt
to assess the impact of stellar activity was to use the
SCALPELS methodology of \cite{collier21}. In summary, SCALPELS attempts to
distinguish dynamically-produced RV variations from activity-induced distortions
on each spectrum's CCF by projecting the
RV time series onto the ten highest variance principle components of the
autocorrelation function of each CCF. The shape changes showed no discernible
trends or periodicity on timescales from 3 days to the duration of the HARPS-N
campaign (i.e. 210 days). We concluded that the effects of stellar activity on
the measured RVs are unmeasurable with our data.

In defiance of the outcome from SCALPELS, we also attempted
to model the weakly correlated RV residuals using an untrained quasi-period GP.
The quasi-periodic covariance kernel is parameterized
by the covariance amplitude $a_{\text{GP}}$, the exponential decay timescale of
active regions $\lambda_{\text{GP}}$, the coherence $\Gamma_{\text{GP}}$, and the
periodic timescale $P_{\text{GP}}$, often related to \prot{} or one of its
low-order harmonics. These four GP hyperparameters are appended to the set of
model parameters, thus resulting in a total of 20 model parameters. Our GP
implementation methodology is standard and has been outlined in detail in
previous work \citep{cloutier19c,cloutier20b}.
We have no prior constraints on the GP hyperparameters from a training set
because no available activity-sensitive time series shows evidence for stellar
activity. We attempted two flavors of GP modeling: firstly with no prior on any
of the GP hyperparameters and secondly with a prior on $P_{\rm GP}$ based on the
estimated \prot{} $=77^{+26}_{-20}$ days for M dwarf rotation-activity relations.
The results
from both MCMCs yielded no constraints on the remaining GP hyperparameters and
more importantly, resulted in measurements of the planet's semiamplitude that
were consistent with zero. We conclude that the non-deterministic nature of the
untrained GP has too much flexibility and effectively absorbs the planetary
signal. We therefore default to the results from our fiducial model for the
remainder of this study.

\section{Discussion} \label{sect:discussion}
\subsection{Fundamental Planetary Parameters} \label{sect:mr}
From our global light curve plus RV analysis, we find that \name{} b has an
orbital period of $P = 0.989343\pm 0.000015$ days.
Using the stellar parameters from
Table~\ref{tab:star}, this corresponds to a semimajor axis of
$a=0.01545\pm 0.00014$ au and an
insolation flux of $F=121^{+12}_{-11} F_{\oplus}$. Although the tidal
quality factors $Q$ for Super-Earths and sub-Neptunes are largely unknown
\citep{morley17b,puranam18}, for a range of plausible $Q$ factors encompassing
the Earth \citep[$Q_{\oplus} \sim 10$;][]{murray99}, to Uranus and Neptune
\citep[$Q\sim 10^4$;][]{tittemore90,zhang08}, \name{} b's ultra-short period
results in a tidal circularization timescale of $<3$ Myrs. Such a short
circularization timescale strongly suggests that the orbit of \name{} b is
circularized. The corresponding equilibrium dayside temperature of \name{} b is
$T_{\text{eq,day}}=1307\pm 30$ K assuming zero albedo. If we assume efficient heat
redistribution around to the nightside, then the zero-albedo equilibrium
temperature becomes $T_{\text{eq}}=924\pm 22$ K.

\begin{figure}
  \centering
  \includegraphics[width=\hsize]{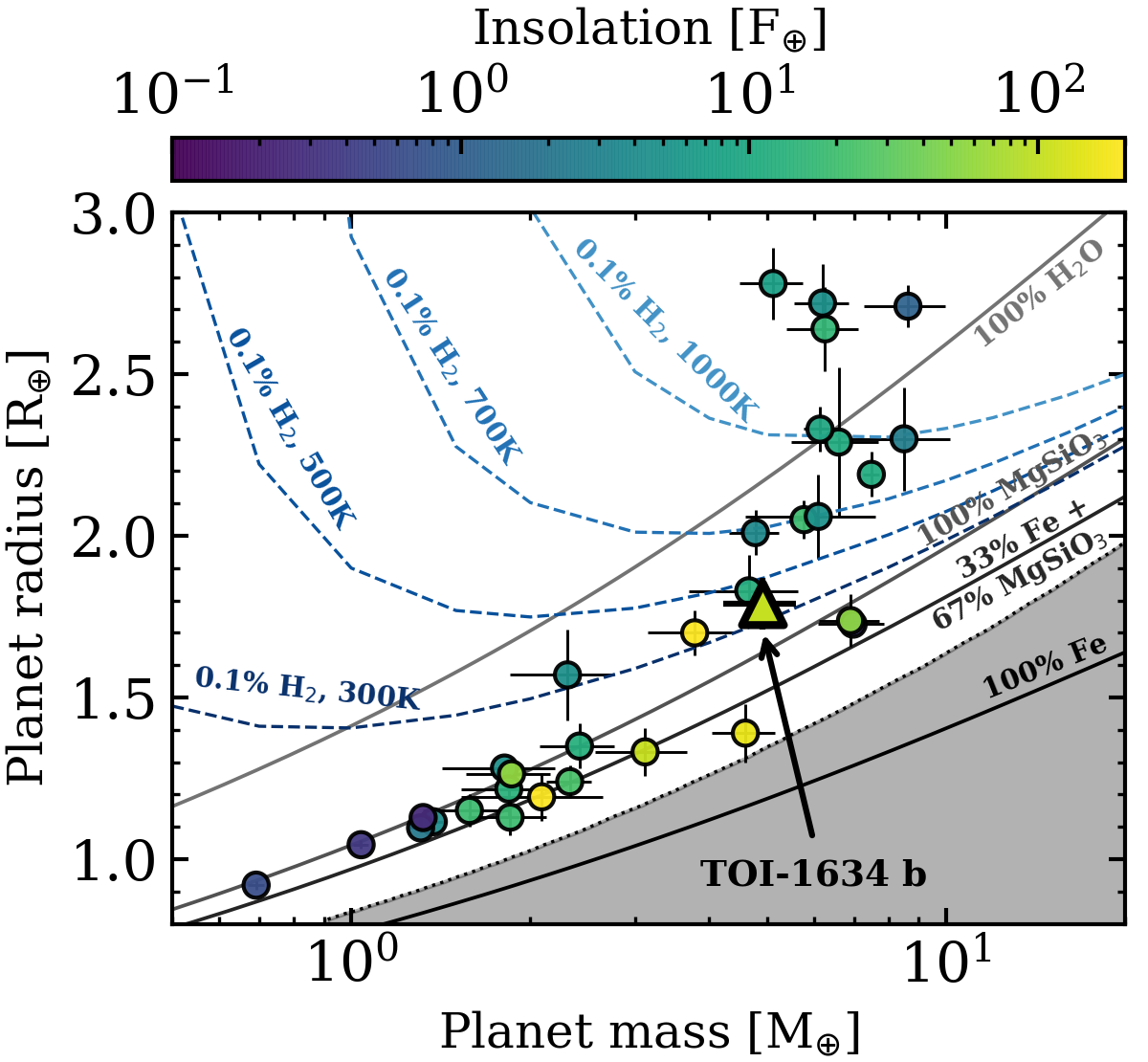}
  \caption{Mass-radius diagram for small planets transiting M dwarfs and with
    precisely measured masses of $\geq 3\sigma$. \name{} b is depicted by the
    lone triangle marker.
    The solid curves are illustrative interior structure models of 100\% water,
    100\% magnesium silicate rock, 33\% iron plus 67\% rock (i.e. Earth-like),
    and 100\% iron \citep{zeng13}. The dashed curves depict models of enveloped
    terrestrials consisting of an Earth-like core enveloped in H$_2$ gas with
    a 1\% envelope mass fraction over a range of equilibrium temperatures.
    The dashed curve bounds the forbidden shaded region according to models of
    maximum collisional mantle stripping by giant impacts \citep{marcus10}.}
  \label{fig:mr}
\end{figure}

We also measure the radius and mass of \name{} b to be
$r_p=$ \rplanet{} \Rearth{} and $m_p=$ \mplanet{} \Mearth{.}
These values correspond to $22\sigma$ and $7\sigma$
detections, respectively. Combining these values gives a $4.7\sigma$ bulk
density measurement of $\rho_p=4.7^{+1.0}_{-0.9}$ g cm$^{-3}$. Figure~\ref{fig:mr}
compares the mass and radius
of \name{} b to the current population of small M dwarf planets with masses
measured to better than $3\sigma$. \name{} b is under-dense compared to an
Earth-like composition planet of the same mass and is inconsistent
with an Earth-like composition at $5.9\sigma$. As such, \name{} b could
belong to the population of enveloped terrestrials whose cores resemble that
of the Earth but also require an extended gaseous envelope to explain their
masses and radii. Assuming an Earth-like planetary core surrounded by a H/He
envelope with solar-metallicity ($\mu=2.35$), whose envelope structure is
described by the semi-analytic radiative-convective model from \cite{owen17},
we find that \name{} b would only
require an envelope mass fraction of $X_{\rm env}=0.30^{+0.19}_{-0.17}$\% to
explain its mass and radius. Here the uncertainties on $X_{\rm env}$ arise from
sampling the marginalized posteriors of $m_p$, $r_p$, and $T_{\rm eq}$.
However, such an extended H/He envelope at 121 times Earth insolation is highly
susceptible to thermally-driven hydrodynamic escape \citep{lopez17}, which
makes \name{} unlikely to be an enveloped terrestrial. Another possibility is
that \name{} b formed beyond the ice line and has retained a volatile-rich
composition \citep{raymond08} with a high mean molecular weight
atmosphere that may be resistant to hydrodynamic escape \citep{lopez17}.
Although we cannot rule out this possibility with our data, a
volatile-rich composition is generally disfavored at the population level as
forward modeling of the radius valley has revealed that the location of the
radius valley strongly favors a smoothly-varying (i.e. not bimodal)
distribution of  
underlying core masses, whose compositions are Earth-like rather than iron or
water-rich \citep{owen17,wu19,gupta19,rogers20}.

Alternatively, the fact that \name{} b appears to be under-dense relative to
an Earth-like composition may be explained by a rocky composition that is
enhanced in Ca and Al-rich minerals rather than the typical Earth-like rocky
compounds of magnesium silicates and iron \citep{dorn19}. At temperatures
exceeding 1200 K within the mid-plane of the protoplanetary disk, the
condensation fraction of Ca and Al is greater than that of Mg, Si, and Fe,
which would provide more solid Ca and Al-rich material from which rocky planets
could form. As such, if \name{} b formed in situ, it could belong to an
alternative class of Super-Earths whose rocky interior compositions differ
significantly from the Earth and the majority of Super-Earths.

Among the M dwarf planets depicted in Figure~\ref{fig:mr} that are under-dense
relative to an Earth-like composition (denoted sub-Neptunes for simplicity), all
of which are larger than 1.7 \Rearth{,} \name{} b is fairly unique in that the
insolation it receives is uncharacteristically high. With an insolation flux of
$F=121^{+12}_{-11} F_{\oplus}$, \name{} b is the second most highly irradiated
sub-Neptune orbiting an M dwarf
\citep[TOI-1685 b receives an insolation flux of $217\: F_{\oplus}$;][]{bluhm21}.
This fact makes \name{} b a somewhat uniquely accessible sub-Neptune for
atmospheric characterization. The physical implications of a high equilibrium
temperature on a sub-Neptune will have to wait for such
observations (see Section~\ref{sect:atm}).
Other examples of well-studied USP sub-Neptunes around FGK stars 
include 55 Cnc e \citep{bourrier18} and WASP-47 e
\citep{vanderburg17}. The exact cause of these peculiar under-dense planets is
unknown but it
has been noted that 55 Cnc and WASP-47 are the most metal-rich stars among
small USP planet hosts ([Fe/H]$_{\text{55 Cnc}}=0.35$ dex,
[Fe/H]$_{\text{WASP-47}}=0.38$ dex; \citealt{dai19}) and
they are the only known systems to contain both a small USP planet and a
close-in giant planet, the presence of which can influence icy pebble drift and
thus the water inventory of the inner disk \citep{bitsch21}.
For comparison, \name{} also appears to be somewhat
metal-rich ([Fe/H]$=0.23^{+0.07}_{-0.08}$ dex) but our RV analysis does not
provide any evidence for an outer giant planet.
Further investigations of these features, and the possiblity that these
USP planets are representative of a new class of Ca and Al-rich Super-Earths,
may provide clues of possible evolutionary pathways that are able to produce
sub-Neptune USP planets.

\subsection{Implications for the emergence of the radius valley around early M dwarfs}
A variety of physical mechanisms have been proposed to explain the emergence of
the radius valley. These include models of thermally-driven
atmospheric mass loss such as photoevaporation: hydrodynamic escape driven by
stellar XUV heating \citep{owen13,jin14,lopez14,chen16,owen17,jin18,lopez18},
and core-powered mass loss: atmospheric heating and escape
driven by the planet's own cooling luminosity
\citep{ginzburg18,gupta19,gupta20}. Conversely, the radius valley has 
also been proposed as a natural outcome of the formation of rocky Super-Earths
and enveloped terrestrials from a gas-poor (but not gas-depleted) environment,
without the need to invoke any subsequent atmospheric escape \citep{lee21}.
When parameterizing the slope of the radius valley via
$r_{p,\text{valley}}\propto P^{\beta}$, each of the photoevaporation, core-powered
mass loss, and gas-poor formation models predict
that $\beta \in [-0.15,-0.09]$ \citep{lopez18,gupta20,lee21}. Whereas, if
enveloped terrestrials form within the first few Myrs when the gaseous disk is
still present, and terrestrial planet formation proceeds at late times after
the dissipation of the gaseous disk in a gas-depleted environment, then the
period-dependence of the radius valley is expected to exhibit the opposite sign
\citep[$\beta = 0.11$;][]{lopez18}.

\begin{figure*}
  \centering
  \includegraphics[width=.9\hsize]{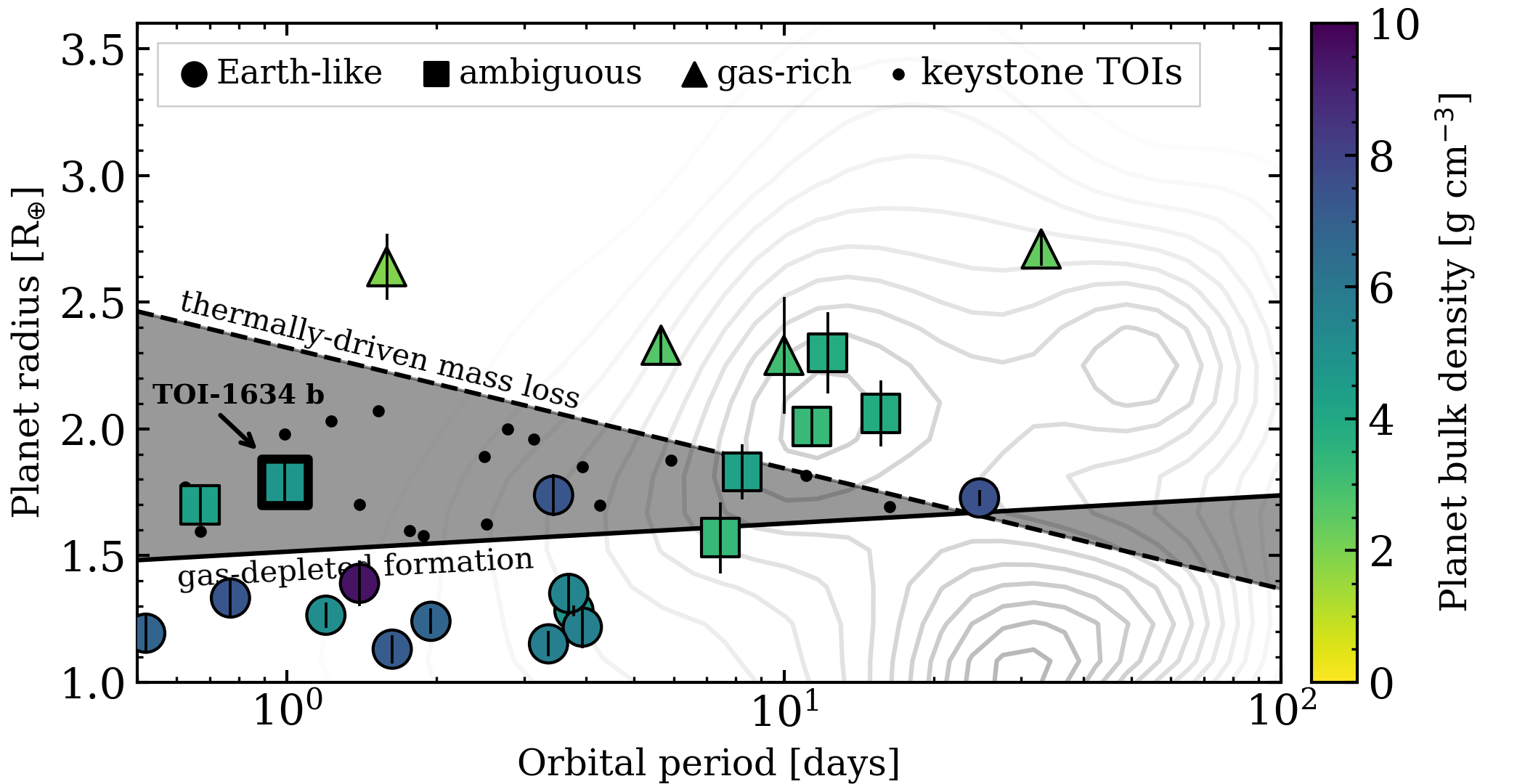}
  \caption{Period-radius diagram for small planets transiting M dwarfs and with
    precisely measured RV masses of $\geq 3\sigma$. The gray contours depict 
    the planetary occurrence rates around low mass stars from Kepler and K2 
    \citep{cloutier20}. The dashed
    and solid lines depict model predictions of the location of the M dwarf
    radius valley from thermally-driven mass loss and from gas-depleted
    formation, respectively. The intermediate shaded regions host the so-called
    keystone planets like the newly discovered \name{} b. The marker shapes
    depict planets whose bulk compositions have been determined to be Earth-like
    (circles), gas-rich (triangles), or ambiguous (see text for possible
    explanations; squares). The colorbar
    highlights each planet's bulk density. The dots depict keystone TOIs that
    have yet to be vetted as validated planets or false positives.}
  \label{fig:radval}
\end{figure*}

The radius valley around Sun-like stars with \teff{} $>4700$ K has been
well-characterized with both Kepler and K2
\citep[e.g.][]{fulton17,vaneylen18,fulton18,martinez19,zink20} and measurements
of $\beta$ take on values $\in [-0.11,-0.09]$ \citep{vaneylen18,martinez19}.
Thus, a thermally-driven mass loss or gas-poor formation model is favored in
this stellar mass regime. However, around lower mass mid-K
to mid-M dwarfs, there is tentative evidence that $\beta$ flattens and becomes
consistent with predictions from gas-depleted formation
\citep[$\beta = 0.06\pm 0.02$;][]{cloutier20}. This suggests that
gas-depleted formation, similar to the suspected formation of the inner solar
system, might begin to dominate the close-in planet population around
M dwarfs. The distinct slopes of the radius valley's period dependence
naturally carve out a region of the orbital period-planet radius parameter space
within which the models make opposing predictions as to whether any planets
located therein should have a rocky Earth-like composition or instead be 
enveloped in H/He gas (Figure~\ref{fig:radval}).
We refer to these radius valley planets around M dwarfs
as keystone planets as they can be used to directly rule out certain models
from precise mass and radius measurements.
We note however that not all keystone planets are equally useful for
constraining model applicability as, for example, keystone planets with
orbital periods between 10-40 days can be consistent with
both models given typical uncertainties on their planetary radii.

With its orbital period of 0.989 days and its size of \rplanet{}
\Rearth{,} \name{} b sits inside of the M dwarf radius valley between
the model predicted slopes (Figure~\ref{fig:radval}). Figure~\ref{fig:radval}
also features the subset of M dwarf planets from Figure~\ref{fig:mr} with
precise RV masses. Planets are classified based on their bulk compositions
inferred from their masses and radii. Earth-like planets are defined as those
consistent with an Earth-like compositional curve, gas-rich planets cannot be
explained by even 100\% water composition and require an extended
H/He envelope, and the intermediate planets we broadly classify as
``ambiguous'' given that they may be explained by a variety of compositions
including a H/He envelope, a volatile-rich composition, or a rocky composition
that is enhanced in Ca and Al. Our analysis revealed that \name{} b is
inconsistent with an Earth-like composition at $5.9\sigma$\footnote{Similarly,
  the mass of \name{} b is inconsistent with a 100\% MgSiO$_3$ composition at
  $2.6\sigma$.} and requires an alternative physical interpretation to explain
its mass and radius. As such, we assign \name{} b to the ``ambiguous'' category.

Regardless of the true composition of \name{} b,
the inconsistency of \name{} b's mass and radius with an
Earth-like composition indicates that it is not compatible with models of
thermally-driven mass loss of Earth-like cores and may support the
gas-depleted formation model. However, this picture may not be so clear 
because models of thermally-driven mass loss have focused on rocky planets
that have Earth-like compositions \citep[e.g.][]{owen17,gupta19}
and have not considered the possiblity of other types of rocky planets that
are Ca and Al-enriched. But if the gas-depleted formation scenario was
operating in the \name{} system then \name{} b would have formed early
on within the gaseous disk's lifetime and subsequently accreted a primordial
H/He envelope that was not entirely lost to space. However, this presents a
clear ambiguity in
that if \name{} b had accumulated a primordial H/He envelope at its current
location, then such an envelope should have been rapidly lost to thermal escape.
The curious case of \name{} b may be reconciled if it migrated inward to
its current location after the extended XUV active phase
\citep[$0.5-1$ Gyr;][]{shkolnik14,france16} and due to its hosting of a
high mean molecular weight atmosphere. Alternatively, \name{} b may indeed be
a rocky planet that is enhanced in Ca and Al and hence is under-dense relative
to an Earth-like rocky planet \citep{dorn19}. However, each of these scenarios
are speculative as they are presently indistinguishable with the data available.

The compositions of the keystone planets classified as
``ambiguous'' make it
difficult to robustly establish gas-depleted formation as the mechanism
responsible. If we wish to establish the importance of the gas-depleted
formation scenario around M dwarfs, we recommend that RV follow-up campaigns
of transiting planets focus on small, potentially-rocky planets with
orbital periods $\gtrsim 20$ days and very precise radii
\citep[e.g. the Super-Earth LHS 1140 b with $P=24.7$ days;][]{dittmann17a}.
The thermally-driven hydrodynamic escape timescales for
such planets are typically longer than the age of the system such that if
they turn out to be rocky, such compositions cannot be explained by
atmospheric mass loss and those planets are likely to have formed rocky. TESS
has already revealed five such planet candidates\footnote{TOIs 198.01, 203.01,
  256.01, 1266.02, 2094.01, \& 2095.02.} that we recommend be targeted for
detailed follow-up.

The fact still remains that the composition of \name{} b does not
resemble an Earth-like composition at $5.9\sigma$, a result which supports the
notion that thermally-driven mass loss may not dominate the
sculpting of the close-in planet population around M dwarfs with masses less
than or similar to \name{} ($\lesssim 0.5$ \Msun{).} This proposition
is further supported by the composition of the other keystone USP planet
in Figure~\ref{fig:radval}: the sub-Neptune
TOI-1685 b whose host stellar mass is nearly identical to that of \name{}
\citep[$M_{s,1685}=0.495$ \Msun{;}][]{bluhm21}. However, because models
of thermally-driven mass loss have assumed that the underlying cores of
the close-in planet population are Earth-like, if \name{} b is rocky but not
Earth-like (i.e. Ca and Al-enriched), then its mass and radius may still be
consistent with models of thermally-driven mass loss.

\subsection{Prospects for atmospheric characterization} \label{sect:atm}
The mass and radius of \name{} b strongly suggest the presence of a gaseous
envelope, which may be accessible with JWST. 
Given the unusual combination of \name{} b's location in mass-radius space
and its close orbital separation, \name{} b occupies a unique region of the
parameter space wherein it would be particularly interesting to distinguish
between different atmospheric compositions. For example, a H$_2$O-rich
atmosphere would likely be indicative of a substantial initial water reservoir
\citep{schaefer16,kite20} whereas a CO$_2$-rich atmosphere may be produced by a
runaway greenhouse if a significant portion of the planet's water inventory was
photolyzed and lost to space. Given its ultra-short period, \name{} b is an
attractive candidate to distinguish between these atmosphere models via thermal
emission observations as atmospheric signatures are likely to be more easily
accessible than in transmission \citep{morley17}. Emission versus transmission
spectroscopy is also less susceptible to signal attenuation by either
clouds/hazes or high mean molecular atmospheres \citep{millerricci09}, and it
can also be used to probe the atmospheric temperature profile.

Assuming a dayside temperature of $T_{\text{eq,day}}=1307$ K, the analytical
emission spectroscopy metric \citep[ESM;][]{kempton18} for
\name{} is approximately 23. This value illustrates \name{} b's favorability for
emission spectroscopy observations when compared to the ESM values of the 
flagship M dwarf planets LHS 3844 b \citep{vanderspek19}, GJ 1132 b
\citep{berta15}, and TRAPPIST-1 b \citep{gillon17a}, whose ESM
values are 30, 10, and 4, respectively.
However, we note that unlike \name{} b, each of these planets is consistent
with a rocky bulk composition. \cite{malik19} calculated the number of
eclipse observations needed to distinguish between either of the aforementioned
atmospheric scenarios and a clear solar composition atmosphere for the
flagship M dwarf planets LHS 3844 b, GJ 1132 b,
TRAPPIST-1 b. By scaling the \cite{malik19} results for GJ 1132 b to the
expected emission S/N of \name{} b, we estimate that the H$_2$O and CO$_2$-rich
atmospheres could be distinguishable for one another with 2-4 JWST/MIRI
eclipses in its slitless
LRS mode. Similarly, we expect 5 eclipses are required in order to distinguish
between the H$_2$O and solar composition models using NIRSpec/G395M.

For the sake of completeness, we also estimate the number of transit
observations needed to detect transmission features. Assuming an isothermal
temperature profile at the zero-albedo equilibrium temperature
$T_{\text{eq}}=924$ K, the expected depths of transmission features at
two scale heights \citep{stevenson16,fu17} in a solar composition,
H$_2$O-dominated, or CO$_2$-dominated atmosphere are 102, 13, and 5 ppm,
respectively. We simulate NIRISS/SOSS ($0.8-2.8 \mu$m) and NIRSpec/G395M
($2.8-5.2 \mu$m) observations using \texttt{PandExo} \citep{batalha17} and
find that only features in a clear solar composition atmosphere would be
detectable at $\geq 3\sigma$ with fewer than 10 transits. However, a clear
H/He-dominated atmosphere for \name{} b is highly unlikely given that
such an atmosphere is unstable at 121 times Earth's insolation flux
\citep{lopez17}. High altitude
clouds would also be increasingly detrimental to feature detection in
transmission, whereas the presence of clouds may be more easily inferred with
secondary eclipse observations as a high dayside albedo would distinguish clouds
from a bare rocky surface with fewer than 10 visits \citep{mansfield19}.

\subsection{Constraints on additional planets} \label{sect:sens}
\subsubsection{RV sensitivity}
M dwarfs hosting multi-planet systems are ubiquitous. Focusing on Kepler stars
with \teff{} $<4000$ K and \logg{} $>3$, \cite{dressing15a}
found that late K to early M dwarf stars host $2.5\pm 0.2$ planets smaller than
4 \Rearth{} and out to 200 days per star. Similarly, \cite{gaidos16} confirmed
these results over a similar range of planetary radii
(1-4 \Rearth{)} and orbital periods (1.5-180 days): $2.2\pm 0.3$ planets per
early M dwarf. Complimentary studies of M dwarf planet occurrence
rates from RV studies have yielded similar results to those
obtained from the Kepler transit survey \citep{bonfils13,tuomi14}.
Although largely limited by poor counting statistics from the
Kepler mission, preliminary occurrence rate calculations around mid-M dwarfs
from \cite{muirhead15} and \cite{hardegree19} have posited that compact systems
($P<10$ days) of multiple planets are common, and perhaps increasingly so as
the host stars become less massive. Indeed, there have been a number of
apparently single transiting M dwarf systems which later revealed additional
planets following one or both of photometric and RV follow-up (e.g. GJ 357c,d;
\citealt{luque19}, GJ 1132c; \citealt{bonfils18}, GJ 3473c; \citealt{kemmer20},
K2-18c; \citealt{cloutier19a}, LHS 1140c; \citealt{ment19}).

With the existence of one known planet orbiting \name{,} it is reasonable to
think that a second planet may exist but as yet remains undetected
because of its small size, long orbital period, or because its orbit is not in a
transiting configuration. Here we place limits on hypothetical planets around
\name{} given our RV time series. Specifically, we compute the
RV sensitivity to planets around \name{} as a function of planet mass 
and orbital period via a set of injection-recovery tests. We run a Monte Carlo
simulation of $10^4$ realizations by injecting synthetic Keplerian signals into
the residuals of the HARPS-N RV time series after the removal of the
MAP solution for \name{} b. In each realization, we simulate a single planet.
Planet masses and orbital periods are sampled uniformly in log-space with the
following bounds: 0.1-20 \Mearth{} and 1-200 days.
Orbital phases are sampled uniformly from $\mathcal{U}(0,2\pi)$.
We sample the orbital inclination from the Gaussian distribution
$\mathcal{N}(i_b,\sigma_i)$, where $i_b=88.2^{\circ}$ and the dispersion in
mutual inclinations of $\sigma_i=2^{\circ}$
follows from studies of multi-planet M dwarf systems \citep{ballard16}. The
stellar mass is also sampled from its posterior and is used to calculate
the corresponding RV semiamplitude assuming a circular orbit. We then inject the
synthetic Keplerian signal into the RV residuals, thus preserving any residual
noise from systematics or uncorrected stellar activity. The
individual measurement uncertainties and timestamps are left unchanged.

\begin{figure}
  \centering
  \includegraphics[width=\hsize]{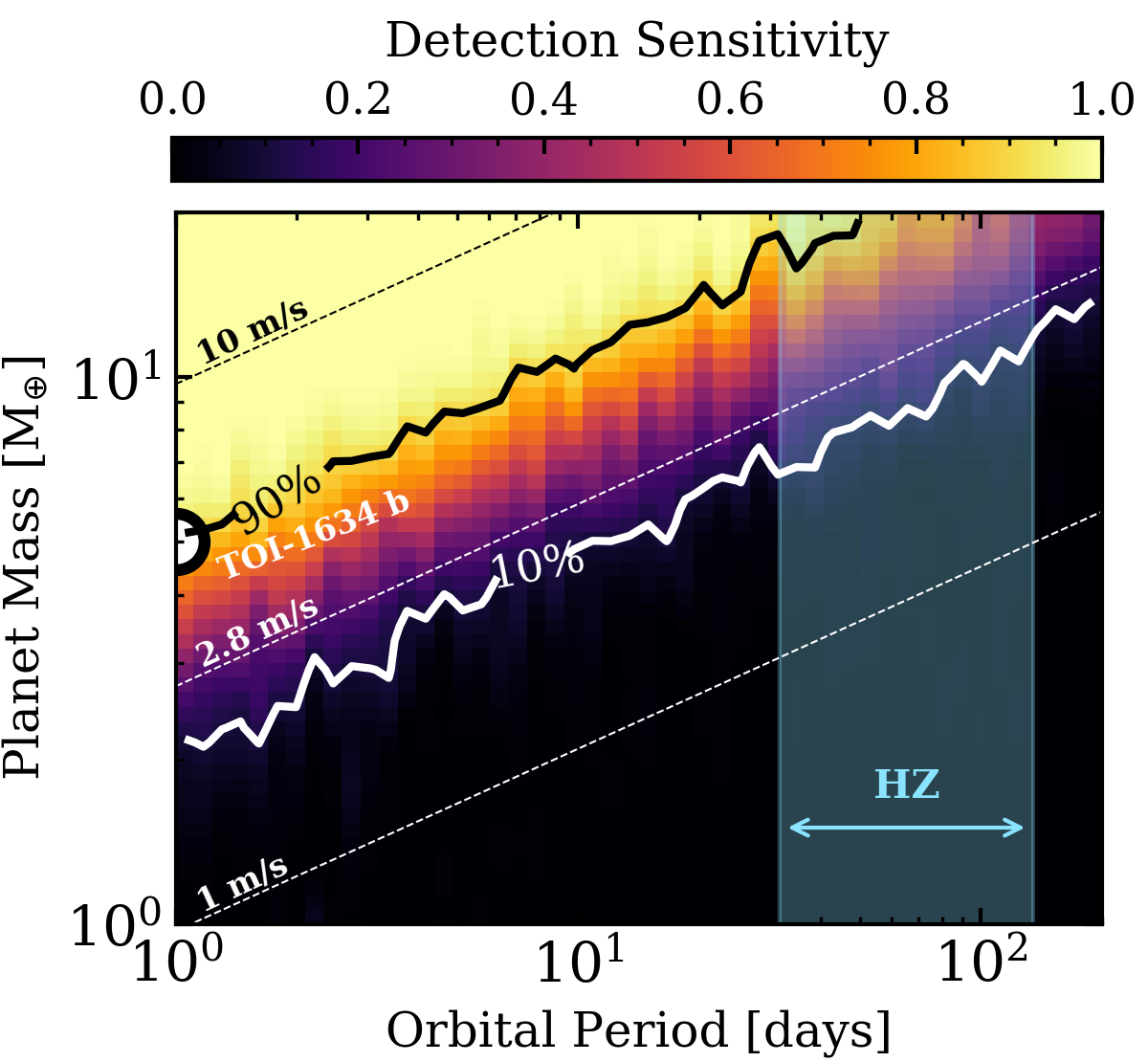}
  \caption{RV detection sensitivity to planets orbiting \name{} as a function
    of planet mass and orbital period. The solid line contours
    highlight the 10\% and 90\% sensitivity levels. The thin dashed lines
    represent lines of constant semiamplitude with illustrative values equal
    to 1 \mps{,} the median dispersion in our HARPS-N time series
    (2.8 \mps{)}, 
    and 10 \mps{.} The circle marker highlights \name{} b. The vertical
    shaded region spans the habitable zone of \name{} whose inner and outer
    edges are defined by the recent Venus and early Mars boundaries
    \citep{kopparapu13}.}
  \label{fig:sens}
\end{figure}

We attempt to recover injected planets following a two-step process. The search
for non-transiting planets in RV time series does not have the benefit of
a-priori knowledge of the planet's period and phase. Instead, probable
signals must show a prominent periodic signal in the GLS periodogram with a FAP
$\leq 1$\%\footnote{However, we note that this need not be the case for
  massive outer companions, which can induce detectable long-term trends without
  a prominent signal in the GLS at its orbital frequency.}. Secondly,
the six-parameter Keplerian model must
be strongly favored over the null hypothesis (i.e. a flat line). For the
purpose of model comparison, we adopt the Bayesian Information Criterion
BIC $= 2\ln{\mathcal{L}} + \nu \ln{N}$, where $\mathcal{L}$ is the likelihood of
the RV data given the assumed model, $\nu$ is the number of model parameters,
and $N=32$ is the number of RV measurements. Taken together,
we claim the successful recovery of an injected planet if and only if the
GLS periodogram power of the largest periodic signal within 10\% of the
injected period has FAP $\leq 1$\%
and the BIC value of the Keplerian model is greater than ten times
the BIC of the null hypothesis. The sensitivity of our RV dataset is defined as
the ratio of number of recovered planets over the number of injected planets
and is depicted in Figure~\ref{fig:sens}.

Unsurprisingly, we find that the mass and orbital period of \name{} b
lie within the region where our RV sensitivity is high (i.e. 90\%).
Figure~\ref{fig:sens} also reveals that at an orbital period of 1 day, we are
sensitive to approximately 50\% of planets at 3 \Mearth{} and to all planets
$\gtrsim 5$ \Mearth{.} Within 10 days, we are sensitive to all planets
$\geq 10$ \Mearth{.} Our sensitivity to the majority of terrestrial planets
($m_p \lesssim 5$ \Mearth{)} at orbital periods $>3$ days is relatively poor due
to the large RV dispersion and the modest number of measurements ($N=32$). 
If we adopt the empirical recent Venus and early Mars habitable zone (HZ)
limits from \cite{kopparapu13} (i.e. 30-125 days), we find that we are only
sensitive to very massive HZ planets ($>15$ \Mearth{).} Unfortunately, any such
planet would be
an improbable candidate for habitable conditions given that the most massive M
dwarf planets with Earth-like bulk compositions are less massive than 7-8
\Mearth{} ($m_{p,\text{LHS 1140 b}}=7.0\pm 0.9$ \Mearth{;} \citealt{ment19},
$m_{p,\text{TOI-1235 b}}=6.9\pm 0.8$ \Mearth{;} \citealt{cloutier20c}) such that
any planet whose mass exceeds 15 \Mearth{} would have to host a massive
H/He envelope, thus rendering its surface uninhabitable.

\subsubsection{Search for transit timing variations}
We also conduct a search for transit timing variations (TTVs)
by fitting the individual TESS transits in the detrended \texttt{PDCSAP} light
curve and then select high S/N seeing-limited transits from
Figure~\ref{fig:sg1}. We apply a transit model with all
model parameters fixed other than the time of mid-transit. We fit the 20 TESS
and 11 seeing-limited transits separately and achieve typical photometric
precisions on the individual transit times of 4.4 and 7.5 minutes,
respectively. We find that the deviation of the individual
transit times from a linear ephemeris is consistent with a flat line and shows
a low rms of approximately 30 seconds in the TESS transits, which are of
comparatively higher quality. We conclude that \name{} b shows no
evidence for TTVs.

\subsection{An Independent Analysis of the \name{} System}
Through the international TFOP collaboration, multiple RV teams began
independent follow-up campaigns to characterize the mass of the planet candidate
TOI-1634.01. Our work based on HARPS-N data presented herein represents one
such effort but we acknowledge a second RV analysis of this system, which was
conducted independently of our own \citep{hirano21}. The submissions of these
works were coordinated between the two groups but their respective data,
analyses, and manuscripts were kept intentionally separate.

\cite{hirano21} presented the mass characterzation of two USP planets,
including \name{} b, using infrared RV measurements taken with the IRD
spectrograph at Suburu \citep{tamura12}. Their resulting RV
semiamplitude is discrepant from ours at $5\sigma$ as they measure a larger
RV semiamplitude of $K=10.8\pm 1.0$ \mps{} and show that
\name{} b is likely consistent with an Earth-like composition. Similarly, the
second USP planet presented in \cite{hirano21} (TOI-1685 b) was also found to
be consistent with an Earth-like composition whereas this planet was
previously shown
to be under-dense relative to an Earth-like composition using CARMENES RV
measurements \citep{bluhm21}. The similar analyses conducted in this work,
in \cite{hirano21}, and in \cite{bluhm21} suggest that the differences in
the IRD results compared to HARPS-N for \name{} b and compared to CARMENES for
TOI-1685 b, are derived from the IRD data and not from issues with any one
group's analysis. The exact cause of these discrepancies is currently unknown
and their resolution is left as a future exercise.

\section{Summary and Conclusions} \label{sect:summary}
We presented the discovery of \name{} b, an ultra-short period keystone planet
orbiting an M2 dwarf, which sits within radius valley. Keystone planets
are useful because knowledge of their bulk composition may be used to
distinguish between radius valley emergence models of thermally-driven mass loss
and gas-depleted formation. Our work has produced the following main findings:

\begin{enumerate}
\item \name{} b is a sub-Neptune USP planet
  with $P=0.989343\pm 0.000015$ days, $r_p=$ \rplanet{} \Rearth{,} and
  $m_p=$ \mplanet{} \Mearth{.} The mass and radius of \name{} b are
  inconsistent with an Earth-like composition at $5.9\sigma$.
\item The composition of \name{} b deviates from expections
  from the close-in planet population and may be explained by either a
  volatile-rich layer with a high mean molecular weight atmosphere that is
  resistant to atmospheric loss, or by a rocky composition that is Ca and
  Al-enriched and consequently under-dense relative to the Earth.  
\item The bulk composition of \name{} b is inconsistent with models of
  thermally-driven mass loss (i.e. photoevaporation and core-powered mass loss)
  and with gas-poor formation. Instead, \name{} b appears to support the
  gas-depleted formation model and would suggest that this formation mechanism
  may start to dominate the close-in planet population around M dwarfs with
  masses $\lesssim 0.5$ \Msun{} if indeed \name{} b is not rocky.
\item Emission spectroscopy observations will help to establish the chemical
  and physical properties that make the atmosphere of \name{} b resistant to
  hydrodynamic escape. Atmospheric models of solar composition,
  H$_2$O-dominated, and CO$_2$-dominated may be distinguished with 2-5 eclipse
  observations with JWST/MIRI or JWST/NIRSpec. 
\item Upon evaluating our RV sensitivity to additional planets, we are able to
  rule out terrestrial-mass planets more massive than 5 \Mearth{} within
  2-3 days and planets
  $>15$ \Mearth{} within the star's habitable zone between 30-125 days.
\end{enumerate}

The inconsistency of the mass and radius of \name{} b with an Earth-like
composition suggests that the gas-depleted formation scenario is favored over
thermally-driven mass loss to explain its mass and radius. However, the
unknown underlying composition of \name{} b makes this statement nonrobust.
To determine the applicability of gas-depleted formation around M dwarfs, we
advocate for the mass characterization of small planets with periods
$\gtrsim 20$ days. If these planets are determined to be predominantly rocky,
that would support the gas-depleted formation interpretation because the
mass loss timescales for these planets are too long for thermally-driven mass
loss to explain their rocky compositions.

\acknowledgements
R.C. is supported by a grant from the National Aeronautics and Space
Administration in support of the TESS science mission.

C.D.D acknowledges support from the Hellman Fellows Fund, the Alfred P. Sloan 
Foundation, the David \& Lucile Packard Foundation, and the NASA Exoplanets 
Research Program (XRP) through grant 80NSSC20K0250.

This work is made possible by a grant from the John Templeton Foundation. The
opinions expressed in this publication are those of the authors and do not
necessarily reflect the views of the John Templeton Foundation. This material
is based on work supported by the National Aeronautics and Space Administration
under grant No. 80NSSC18K0476 issued through the XRP Program.

The financial support from the agreement ASI-INAF n.2018-16-HH.0 is gratefully
acknowledged.

Based on observations made with the Italian Telescopio Nazionale Galileo (TNG)
operated by the Fundaci\'on Galileo Galilei (FGG) of the Istituto Nazionale di
Astrofisica (INAF) at the Observatorio del Roque de los Muchachos (La Palma,
Canary Islands, Spain).

The HARPS-N project has been funded by the Prodex Program of the Swiss Space
Office (SSO), the Harvard University Origins of Life Initiative (HUOLI), the
Scottish Universities Physics Alliance (SUPA), the University of Geneva, the
Smithsonian Astrophysical Observatory (SAO), the Italian National Astrophysical
Institute (INAF), the University of St Andrews, Queens University Belfast, and
the University of Edinburgh.

This work has made use of data from the European Space
Agency (ESA) mission Gaia (\url{https://www.cosmos.esa.int/gaia}), processed by
the Gaia Data Processing and Analysis Consortium (DPAC,
\url{https://www.cosmos.esa.int/web/gaia/dpac/consortium}). Funding for the
DPAC has been provided by national institutions, in particular the institutions
participating in the Gaia Multilateral Agreement.

This work makes use of observations from the LCOGT network. Part of the LCOGT
telescope time was granted by NOIRLab through the Mid-Scale Innovations
Program (MSIP). MSIP is funded by NSF.

This article is based on observations made with the MuSCAT2 instrument,
developed by ABC, at Telescopio Carlos Sánchez operated on the island of
Tenerife by the IAC in the Spanish Observatorio del Teide.

This work is partly supported by JSPS KAKENHI Grant Numbers JP18H01265 and
JP18H05439, and JST PRESTO Grant Number JPMJPR1775, and a University Research
Support Grant from the National Astronomical Observatory of Japan (NAOJ).

Resources supporting this work were provided by the NASA High-End Computing
(HEC) Program through the NASA Advanced Supercomputing (NAS) Division at Ames
Research Center for the production of the SPOC data products.

Some of the Observations in the paper made use of the High-Resolution Imaging
instrument $'$Alopeke . $'$Alopeke was funded by the NASA Exoplanet Exploration
Program and built at the NASA Ames Research Center by Steve B. Howell, Nic
Scott, Elliott P. Horch, and Emmett Quigley. $'$Alopeke was mounted on the
Gemini North telescope of the international Gemini Observatory, a program of
NSF's OIR Lab, which is managed by the Association of Universities for Research
in Astronomy (AURA) under a cooperative agreement with the National Science
Foundation on behalf of the Gemini partnership: the National Science
Foundation (United States), National Research Council (Canada), Agencia
Nacional de Investigaci\'on y Desarrollo (Chile), Ministerio de Ciencia,
Tecnolog\'ia e Innovaci\'on (Argentina), Minist\'erio da Ci\^encia, Tecnologia,
Inova\c{c}\~{o}es e Comunica\c{c}\~{o}es (Brazil), and Korea Astronomy and Space
Science Institute (Republic of Korea).

\facilities{TESS, ASAS-SN, TRES, LCOGT, MuSCAT2, OAA, RCO, Gemini/$'$Alopeke, Lick/ShARCS, TNG/HARPS-N.}

\software{\texttt{AstroImageJ} \citep{collins17},
  \texttt{astropy} \citep{astropyi,astropyii},
  \texttt{BANZAI} \citep{mccully18},
  \texttt{batman} \citep{kreidberg15},
  \texttt{BGLS} \citep{mortier15},
  \texttt{celerite} \citep{foremanmackey17},
  \texttt{emcee} \citep{foremanmackey13},
  \texttt{exoplanet} \citep{foremanmackey19},
  \texttt{PandExo} \citep{batalha17},
  \texttt{PyMC3} \citep{salvatier16},
  \texttt{scipy} \citep{scipy},
  \texttt{STARRY} \citep{luger19},
  \texttt{Tapir} \citep{jensen13},
  \texttt{TERRA} \citep{anglada12}.}

\restartappendixnumbering
\appendix
\section{Additional Planet Validation Figures}

\begin{figure*}
  \centering
  \includegraphics[width=.48\hsize]{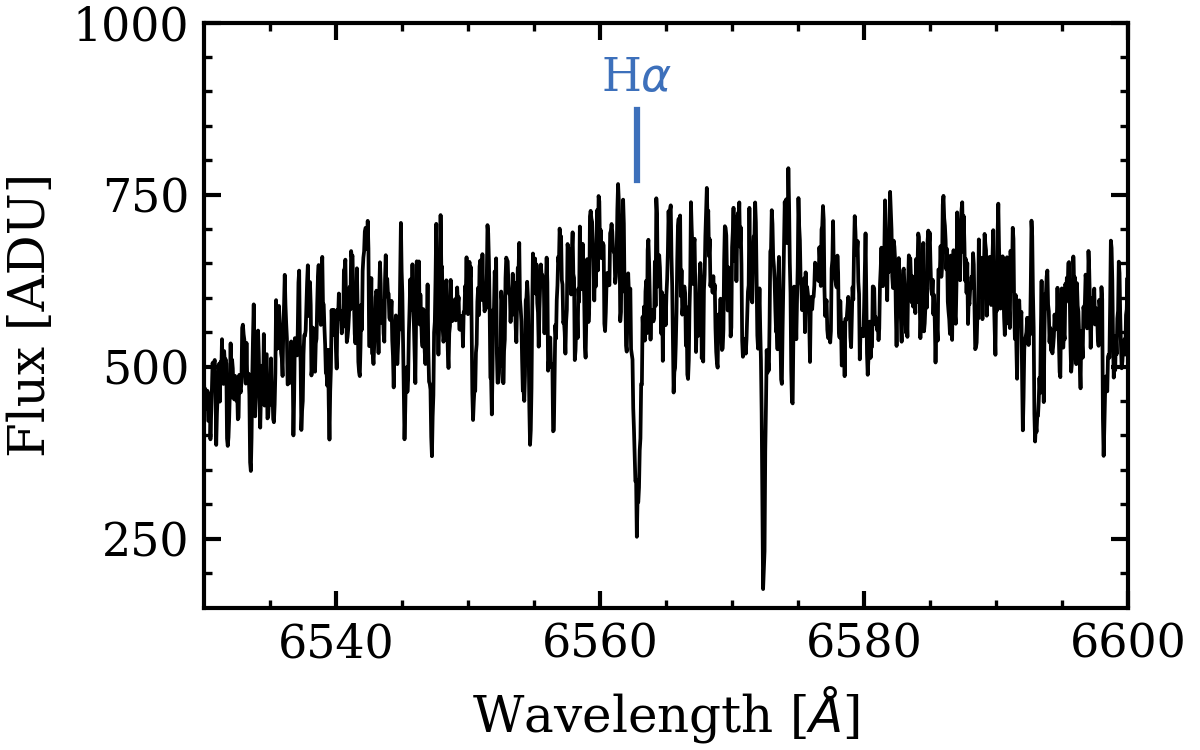}
  \includegraphics[width=.48\hsize]{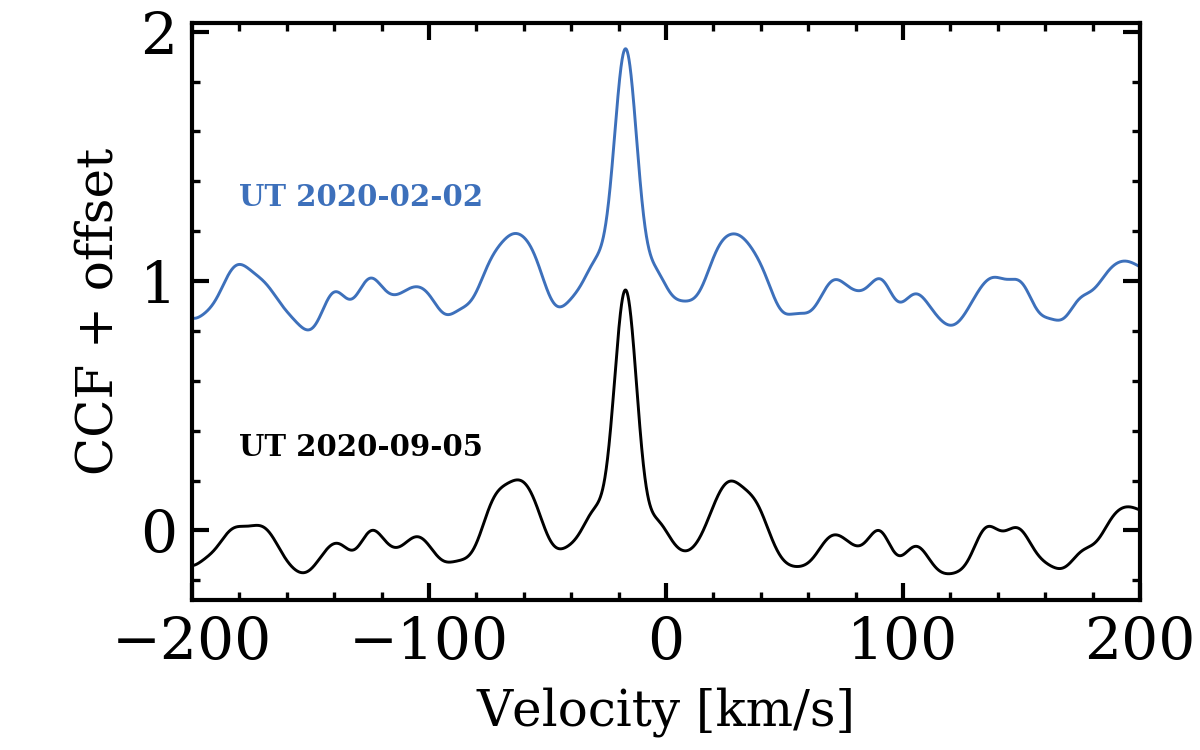}
  \caption{Key takeaways from TRES reconnaissance spectroscopy of \name{.} Left
    panel: $H\alpha$ in absorption indicating a chromospherically
    inactive star. Right panel: the CCF of each TRES epoch reveals a
    single-lined spectrum with no resolved rotational broadening
    (i.e. \vsini{}$<3.4$ \mps{)}.}
  \label{fig:sg2}
\end{figure*}

\begin{figure}
  \centering
  \includegraphics[width=\hsize]{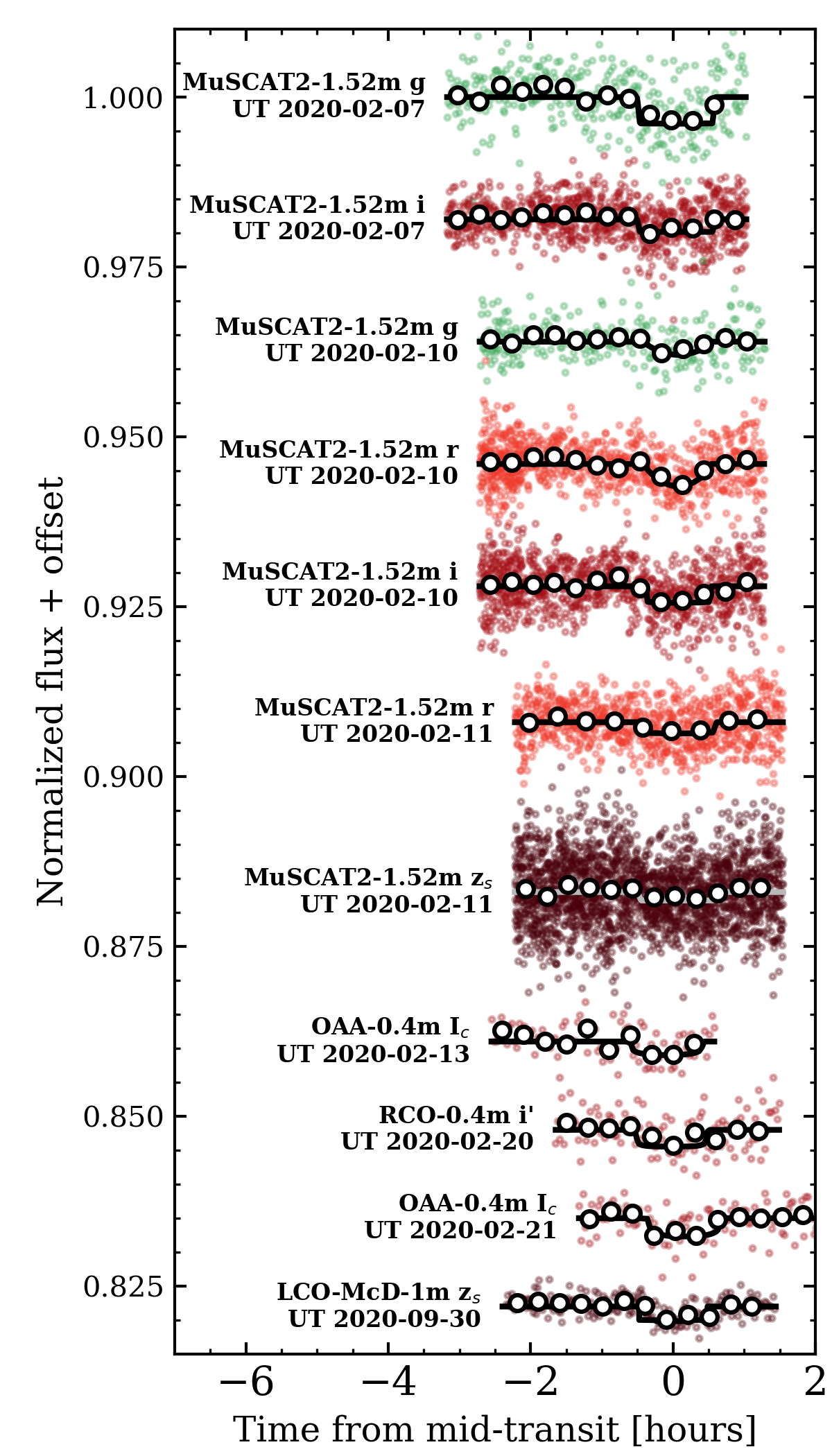}
  \caption{Select seeing-limited photometric light curves of \name{} b transit
    events with particularly low S/N light curves omitted.
    The light curves are vertically offset for clarity. Annotated next
    to each light curve is the observing facility, the passband, and the UT
    observation date.}
  \label{fig:sg1}
\end{figure}

\begin{figure*}
  \centering
  \includegraphics[width=.48\hsize]{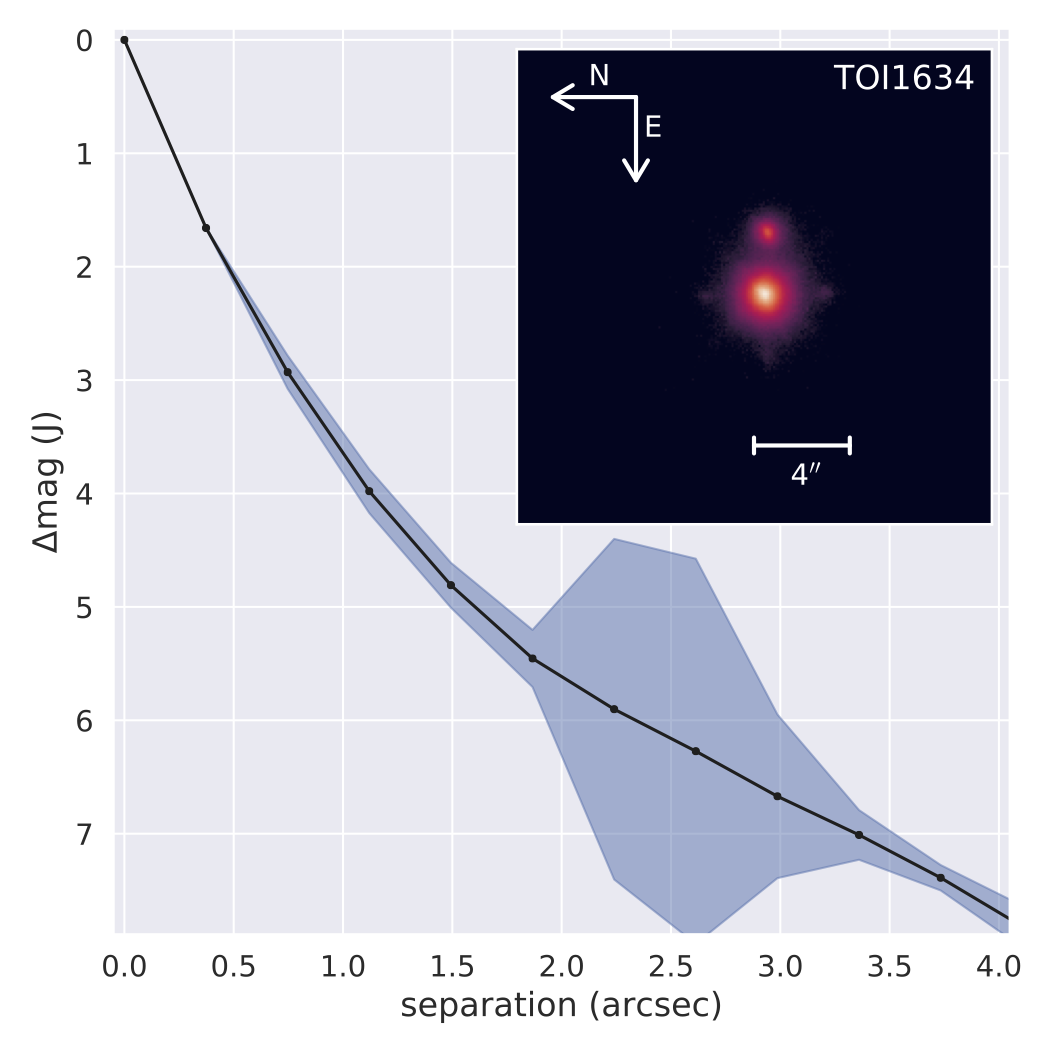}
  \includegraphics[width=.48\hsize]{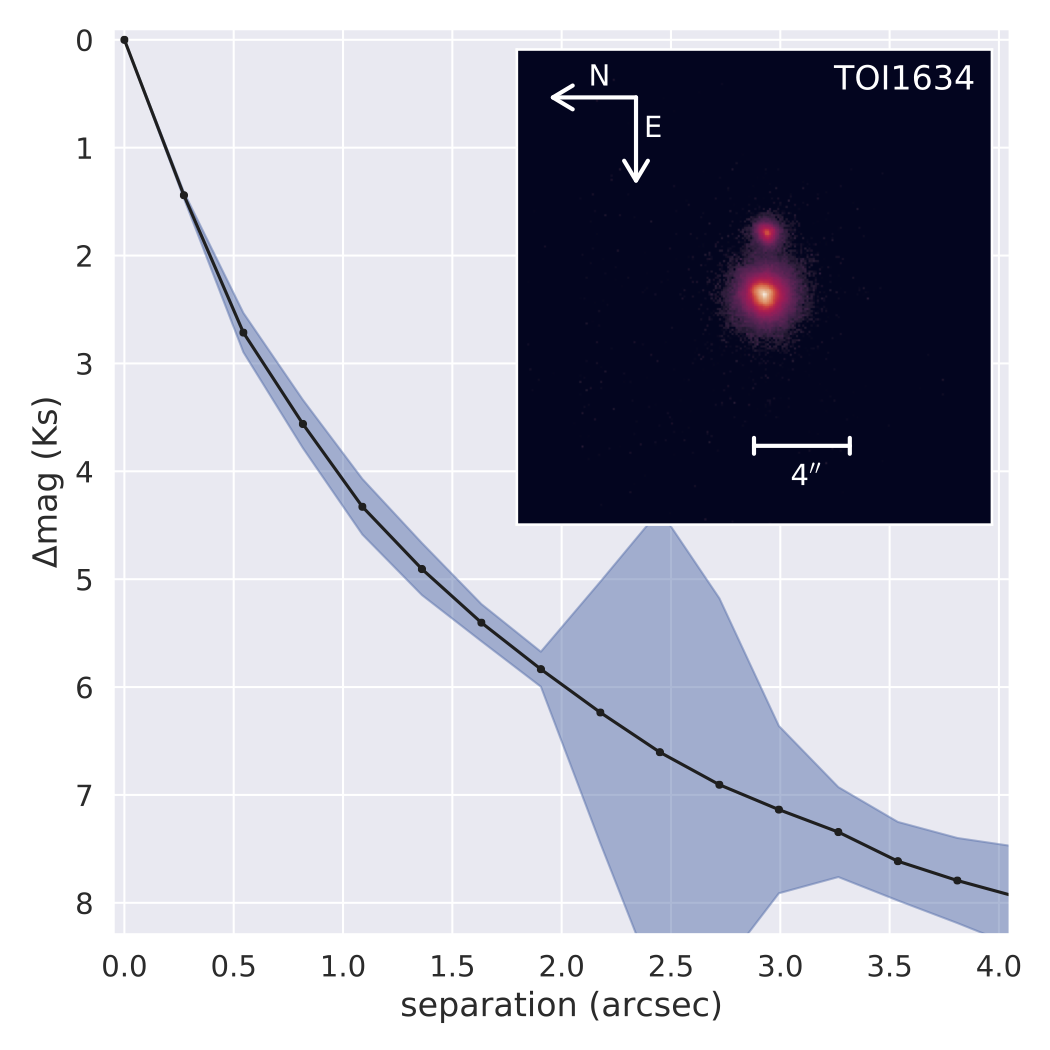}
  \includegraphics[width=.48\hsize]{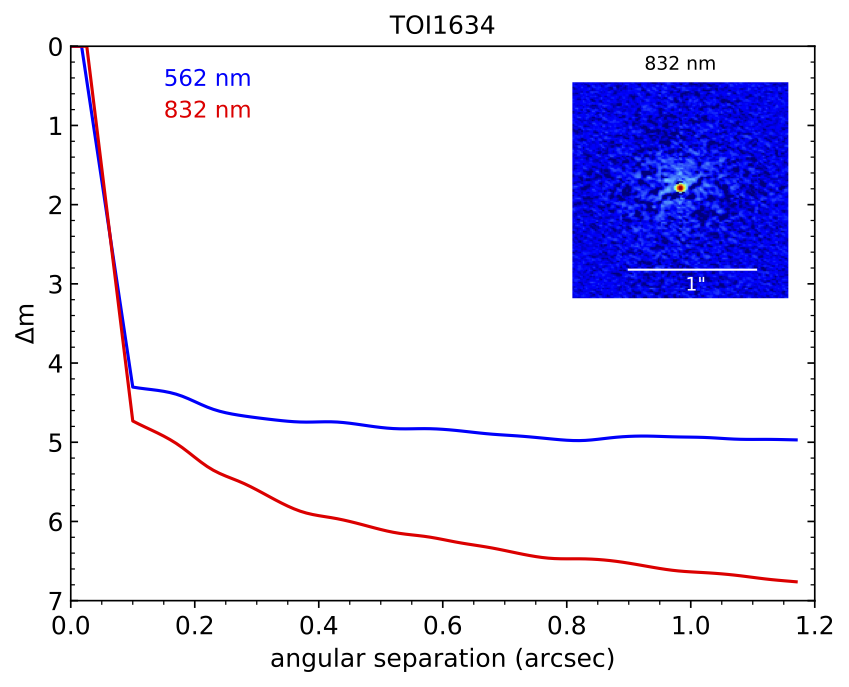}
  \caption{High-resolution imaging observations of \name{.} First and second
    panels: the $J$ (left) and $K_S$-band (right) contrast curves and rms
    errors from ShARCS AO-imaging. The inset in each panel shows the band's
    stacked image. Third panel: $5\sigma$ contrast curves from
    $'$Alopeke speckle imaging
    at 562 nm and 832 nm along with the reconstructed 832 nm image in the
    inset. Other than the known companion located $2 \farcs 69$ west of \name{,}
    no additional companions or background sources are revealed by our high
    resolution imaging.}
  \label{fig:sg3}
\end{figure*}

\bibliographystyle{apj}
\bibliography{refs}

\suppressAffiliationsfalse
\allauthors
\end{document}